\pdfoutput=1
\documentclass[fleqn,usenatbib]{mnras}

\usepackage{textcomp}
\usepackage{lineno}
\usepackage[T1]{fontenc}
\usepackage{url}
\DeclareRobustCommand{\VAN}[3]{#2}
\let\VANthebibliography\thebibliography
\def\thebibliography{\DeclareRobustCommand{\VAN}[3]{##3}\VANthebibliography}

\usepackage{graphicx}	
\usepackage{amsmath}	
\usepackage{amssymb}	
\usepackage{makecell}
\usepackage{multirow}
\usepackage{booktabs}
\usepackage{subfigure} 

\usepackage{ulem}

\title[]{Galaxy Image Classification using Hierarchical Data Learning with Weighted Sampling and Label Smoothing}

\author[X. M et al.]{
Xiaohua Ma$^{1}$,
Xiangru Li$^{1}$\thanks{E-mail: xiangru.li@gmail.com},
Ali Luo$^{2}$,
Jinqu Zhang$^{1}$,
and Hui Li$^{1}$
\\
$^{1}$School of Computer Science, South China Normal University, No. 55 West of Yat-sen Avenue, Guangzhou 510631, China\\
$^{2}$Key Laboratory of Optical Astronomy, National Astronomical Observatories, Chinese Academy of Sciences, Beijing 100012, China
}

\date{Accepted XXX. Received YYY; in original form ZZZ}

\pubyear{2022}

\begin{document}
\label{firstpage}
\pagerange{\pageref{firstpage}--\pageref{lastpage}}
\maketitle

\begin{abstract}
With the development of a series of Galaxy sky surveys in recent years, the observations increased rapidly, which makes the research of machine learning methods for galaxy image recognition a hot topic. Available automatic galaxy image recognition researches are plagued by the large differences in similarity between categories, the imbalance of data between different classes, and the discrepancy between the discrete representation of Galaxy classes and the essentially gradual changes from one morphological class to the adjacent class (DDRGC). These limitations have motivated several astronomers and machine learning experts to design projects with improved galaxy image recognition capabilities. Therefore, this paper proposes a novel learning method, ``Hierarchical Imbalanced data learning with Weighted sampling and Label smoothing" (HIWL). The HIWL consists of three key techniques respectively dealing with the above-mentioned three problems: (1) Designed a hierarchical galaxy classification model based on an efficient backbone network; (2) Utilized a weighted sampling scheme to deal with the imbalance problem; (3) Adopted a label smoothing technique to alleviate the DDRGC problem. We applied this method to galaxy photometric images from the Galaxy Zoo-The Galaxy Challenge, exploring the recognition of completely round smooth, in between smooth, cigar-shaped, edge-on and spiral. The overall classification accuracy  is 96.32\%, and some superiorities of the HIWL are shown based on recall, precision, and F1-Score in comparing with some related works. In addition, we also explored the visualization of the galaxy image features and model attention to understand the foundations of the proposed scheme. 

\end{abstract}




\section{Introduction}

In recent years, industrial and technological advances in society have brought about better observing instruments and techniques, and the astronomical community has ushered in the era of big data. The construction of a series of large telescopes and the implementation of sky survey programmes have enabled the astronomical community to access vast amounts of data. For example, the Sloan Digital Sky Survey \citep[SDSS;][]{york2000sloan}, the Cosmic Evolution Survey \citep[COSMOS;][]{scoville2007cosmic} and the Large Synoptic Survey Telescope \citep[LSST;][]{tyson2002large} have each produced terabytes of data at each stage, and the LSST produces about 20 terabytes of observations every night.

The classification of galaxies is a fundamental and highly significant task in astronomy, and the morphological classification of galaxies from images is a typical part of the relevant data processing. For such a large volume of galaxy observations, it is no longer feasible for manual classification methods. As a result, it is a problem to develop automatic galaxy image classification methods. Particularly, it is a hot topic to design an automatic machine learning scheme for galaxy image classification. Ellipse and spiral are the two classic categories in the Hubble classification scheme of galaxies. However, there are not only a large number of galaxy images in these two broad categories, but also obvious differences in the morphology of various galaxies in each category. For example, there exist evident differences in oblateness between some elliptical galaxies; certain differences on morphology are also shown on some spiral galaxies due to the variation of their directions relative to earth.
A more detailed classification helps us further understand the laws of galaxy evolution and promote the discovery of new galaxies. Therefore, elliptical galaxies were subdivided into completely round smooth, in between smooth and cigar-shaped according to the degree of flattening. And spiral galaxies were subdivided into edge-on and spiral according to the relative direction with the earth.

There have been a number of inspiring explorations and good progress in classifying galaxy images based on traditional machine learning methods, such as decision trees, support vector machines, and ensemble learning. Based on support vector machines, \citet{malek2013vimos} explored the recognition of stars, quasars and galaxies on optical and near-infrared data (u*, g, r, i, z, Ks) in the VIMOS Public Extragalactic Redshift Survey (VIPERS). \citet{sevilla2015effect} applied Boosted Decision Trees (BDTs) method and catalog features to study the recognition of star/galaxy in SDSS-DR9 (SDSS Data Release 9), improving the purity of the galaxy sample. Based on 10 galaxy features and 7941 galaxy sample data, \citet{sreejith2018galaxy} explored the effect of four statistical learning methods in galaxy recognition. Among them, the random forest method not only has the characteristics of simple formula and easy implementation, but also achieves excellent overall accuracy in the five types of galaxy recognition and two types of galaxy recognition problems. \citet{chao2019study} further investigated the star/galaxy recognition problem based on twelve catalog features and the XGBoost (eXtreme Gradient Boosting) method. This study obtained better results than machine learning models such as Adaboost, GBDT, and had the most significant improvement in the recognition of dark source target. \cite{jose2020galaxy} explored the recognition problem of early-type and late-type galaxies using three machine learning methods based on TOELO and COSMOS data. Among them, deep neural network (DNN) not only has higher recognition accuracy than the other two methods, but also can operate with missing data. \citet{reza2021galaxy} explored the effectiveness of five machine learning methods based on SDSS-DR16 data and 62 catalog features for the recognition of spiral galaxies, elliptical galaxies, merging galaxies and stars, and focused on the impact of oversampling and undersampling methods. The results show that the oversampling-based artificial neural networks and the ExtraTrees method are particularly effective. However, these studies rely more or less on manual sample selection or manual feature design and selection. In case of dealing with large amounts of galaxy data, it is not only time-consuming and laborious, but also difficult to guarantee the optimality of the learned features. In addition, the generalization ability of the model obtained by the above studies has much room for improvement in theory.

Deep learning is a branch of the machine learning methods, and has been a hot topic for model research and application exploration in recent years. Preliminary research results show that deep learning techniques have good potential and advantages in terms of automatic learning of effective features and improvement of model generalisation capabilities. \citet{dieleman2015rotation} trained a 7-layer deep convolutional network model for the first time using more than 50,000 galaxy image samples, and finally won the competition named Galaxy Zoo-The Galaxy Challenge with an RMSE of 0.07492. \citet{lukic2016galaxy} explored the influence of network layers and activation functions on galaxy image classification methods of deep learning based on the Galaxy Zoo dataset. The results show that when the number of network layers reaches a certain number, increasing the number of hidden layers not only do not help to reduce the empirical loss, but also lead to overfitting, and the activation function PReLU outperforms ReLU. The ``empirical loss" quantifies the errors over training data in machine learning. \citet{khalifa2018deep} used an 8-layer convolutional neural network model for the EFIGI catalog\citep{baillard2011efigi} to study the recognition of elliptical, spiral and irregular galaxies, and paid special attention to the role of data augmentation in mitigating overfitting. \citet{zhu2019galaxy} studied the recognition of completely round smooth, in between smooth, cigar-shaped, edge-on, and spiral based on Galaxy Zoo data \citep{lintott2011galaxy, willett2013galaxy} and ResNet architecture. Therefore, the research on galaxy recognition problem based on deep learning focuses on the methods of convolutional neural network, which are characterized by translation invariance and certain rotation invariance, as well as excellent feature learning ability.

Although the works of galaxy classification based on deep learning has made good progress on the basis of related research based on traditional machine learning methods, there are still some problems that need to be solved urgently. In the recognition of multi-class galaxy images, generally the samples from most of the two class pairs are easily distinguished from each other, and only a few two-class pair samples are difficult to be discriminated from each other; however, the votes from the former samples unduly reduces the learning attention and recognition ability of the model for the latter. This phenomenon is called the masking effect caused by large differences in similarity between classes. For example, completely round smooth, in between smooth, spiral, and edge-on (or cigar-shaped) galaxies are easily distinguished from each other, while Edge-on and cigar-shaped galaxies are difficult to be discriminated from each other. The large difference in the number of samples belonging to different classes makes the training data of the galaxy classification model imbalanced. Data imbalance can lead to the result that models excessively focus on learning the features of data-dominated classes, while neglecting those of minority classes. The result is that the learned model has poor performance on the minority class. For example, in the work of \citet{zhu2019galaxy}, the model was trained without imbalance treatment of the minority class, and although the final overall validation accuracy reached 95.2\%, the recall of cigar-shaped galaxies was only 58.62\%. The galaxy morphology gradually changes from one class to its neighbor class, whereas their labels, 1, 2, 3, 4 or one-hot codes, are discrete. This difference makes the learned models less effective.

To address these issues, this paper proposed a novel learning method, ``Hierarchical Imbalanced data learning with Weighted sampling and Label smoothing" (HIWL). The HIWL is a hierarchical learning scheme based on the EfficientNet. The hierarchical strategy is to improve the recognition effect of two minority galaxy samples which are difficult to distinguish from each other, by alleviating the masking effect caused by large differences in similarity between classes. The component model EfficientNet has an excellent balance between performance and speed, which is helpful for the efficiency of the HIWL. In the first layer of hierarchical learning, the two classes that are indistinguishable from each other are merged into one combined class treated as a whole. In the second layer, a sub-model is further trained specifically to discriminate the samples from the two minority classes with more similarity with each other. The HIWL sequentially processes a galaxy image: HIWL firstly recognizes the image with sub-model in the first layer; if this image is discriminated as a sample from the combined class, then it will be classified by the sub-model from the second layer. In the preprocessing of the sub-models, HIWL uses a weighted sampling technology and online data augumentation to alleviate the imbalance situation and increase the diversity of samples. 
In the training phase, HIWL uses label smoothing technology to improve the sensitivity to gradual changing characteristics of galaxy morphology. 
The idea is to introduce a smoothing factor into each label, so that the label is no longer an either/or one-hot code, but a soft label indicating the characteristics of gradual changing characteristics on galaxy morphology. 
We applied HIWL to the Galaxy Zoo-the galaxy challenge data to explore its recognition performance of galaxy morphology. 
In order to understand the foundations of the HIWL model, we also explored the visualization of the model features and its attention area.

\section{Data preparation}
\subsection{Sloan Digital Sky Survey}
The Sloan Digital Sky Survey \citep[SDSS;][]{york2000sloan}, a redshift survey project using the 2.5m aperture telescope at Apache Point Observatory in New Mexico, is considered one of the most successful and influential large-scale sky surveys in astronomical history. The project began in 2000 with the aim of observing a quarter of the entire sky, and is now in its fifth phase of operation. During the operation of the project, different versions of photometric images and spectral data will be released in each phase. The latest data released is the last version in the fourth phase, Data Release 17 (SDSS DR17).

\subsection{Galaxy Zoo}
Galaxy Zoo \citep{raddick2009galaxy} is a citizen science project launched by Oxford University, Portsmouth University, Johns Hopkins University and other research institutions. This project invited many volunteers to classify galaxies after a short-term study. The idea is that for each galaxy image prepared in advance, dozens of volunteers are asked to recognize between Spiral galaxies, elliptical galaxies and non-galaxies and to vote. As the result, the class with the highest number of votes would be recorded as the final class. The project classified 34,617,406 galaxy images based on the endeavours of 82,931 volunteers. Galaxy Zoo 2 \citep{willett2013galaxy}, an extension of the above project, elaborates more complex and detailed morphological features such as barred spiral galaxies. The original data of Galaxy Zoo 2 comes from SDSS DR7, the most influential large-scale sky survey project in the astronomical world, and the classification consistency with professional astronomers is more than 90\%. Therefore, this work used this dataset for investigation. Since then, Galaxy Zoo: Hubble and Galaxy Zoo: CANDELS were conducted. These projects provided large-scale, labeled datasets of high-quality galaxy images.

\subsection{Dataset selection}
\label{ssec:dataset selection}
The dataset and classification criteria provided by Galaxy Zoo 2 are used to prepare the training set, validation set and test set in this paper. The dataset is derived from the competition held by the Galaxy Zoo project on Kaggle: Galaxy Zoo-The Galaxy Challenge. The competition dataset consists of 61,578 labeled galaxy images from SDSS DR7. Each image is of three channels and has a size of 424$\times$424. The label of each galaxy image is a 37-dimensional vector, and each component of the label vector describes the statistical information of the answers to one question in the quiz tasks from the volunteers. The classification criteria and the rules for selecting clean samples come from \cite{willett2013galaxy}.

 To make the samples more representative, a clean sample selection procedure is required. For a galaxy image, it is necessary to ensure that no less than 20 volunteers finish no less than 11 judgement tasks in the decision tree diagram. Different tasks correspond to different questions, and there are various numbers of available responses for each question. There are 37 available responses. Therefore, the recognition result of each galaxy diagram is a 37-dimensional vector. Each dimension of this vector records the frequency of a particular response. A galaxy image will be treated as a clean sample if each component of its recognition vector exceeds the corresponding threshold (Table \ref{table:cleandata select}). In Table \ref{table:cleandata select}, we defined an abbreviation for each galaxy class.

\begin{table*}
\caption{\label{table:cleandata select}Clean sample selection rule \citep{willett2013galaxy}. A sample can only be selected as a clean sample when the response frequency of each corresponding task exceeds a given threshold. In particular, the threshold selection criteria for smooth galaxies (CRS, IBS and CSS) are appropriately relaxed to 0.5 (cf. in \citep{zhu2019galaxy})}.

\begin{tabular}{ccccc}
\hline
Classname & Abbreviation & Task & Selection & $N\_samples$ \\ \hline
Completely round smooth & CRS & \begin{tabular}[c]{@{}c@{}}T01\\ T07\end{tabular} & \begin{tabular}[c]{@{}c@{}}\textit{f}\_smooth\textgreater{}0.469\\ \textit{f}\_completely round\textgreater{}0.50\end{tabular} & 8434 \\
\begin{tabular}[c]{@{}c@{}}In between smooth\end{tabular} & IBS & \begin{tabular}[c]{@{}c@{}}T01\\ T07\end{tabular} & \begin{tabular}[c]{@{}c@{}}\textit{f}\_smooth\textgreater{}0.469\\ \textit{f}\_in between\textgreater{}0.50\end{tabular} & 8069 \\
Cigar-shaped smooth & CSS & \begin{tabular}[c]{@{}c@{}}T01\\ T07\end{tabular} & \begin{tabular}[c]{@{}c@{}}\textit{f}\_smooth\textgreater{}0.469\\ \textit{f}\_cigar-shaped\textgreater{}0.50\end{tabular} & 578 \\
Edge-on & EO & \begin{tabular}[c]{@{}c@{}}T01\\ T02\end{tabular} & \begin{tabular}[c]{@{}c@{}}\textit{f}\_features/disk\textgreater{}0.430\\ \textit{f}\_edge-on,yes\textgreater{}0.602\end{tabular} & 3903 \\
Spiral & SPI & \begin{tabular}[c]{@{}c@{}}T01\\ T02\\ T04\end{tabular} & \begin{tabular}[c]{@{}c@{}}\textit{f}\_features/disk\textgreater{}0.430\\ \textit{f}\_edge-on,no\textgreater{}0.715\\ \textit{f}\_spiral,yes\textgreater{}0.619\end{tabular} & 7806 \\ \hline
\end{tabular}

\end{table*}

After the clean sample selection process, 28,790 samples are obtained. Among them, there are 8434 CRS samples, 8069 IBS samples, 578 CSS samples, 3903 EO samples, and 7806 SPI samples. The training set, validation set and test set are established by randomly dividing the samples from each galaxy class in a ratio of 8:1:1. The training set, validation set, and test set contain 23031, 2878 and 2881 samples, respectively. More information can be found in Table \ref{table:data distribution}. It should be noted that the numbers of samples of CRS, IBS and SPI are large and balanced; in comparison, there are fewer EO, and the number of samples for CSS is the smallest which is less than 1/10 of any one of CRS, IBS and SPI. Therefore, there is an imbalanced relationship between CSS and the other four classes of galaxies, which will bring some difficulties to the training of deep learning models, and the detailed solution is shown in Section \ref{ssec:weighted sampling} of this paper.

 \begin{table}
\centering
\caption{\label{table:data distribution}The division of the training set, validation set and test set. The clean sample dataset is selected through Table \ref{table:cleandata select} and divided into a training set, a validation set and a test set in the ratio of 8:1:1. The numbers in this table represent the number of different classes in the corresponding dataset.}
\resizebox{\linewidth}{!}{
\begin{tabular}{ccccccc}
\hline
 & CRS& IBS & CSS& EO & SPI& Total \\ \hline
Clean dataset & 8434 & 8069 & 578 & 3903 & 7806 & 28790 \\
Training set & 6747 & 6455 & 462 & 3122 & 6245 & 23031 \\
Validation set & 843 & 807 & 58 & 390 & 780 & 2878 \\
Test set & 844 & 807 & 58 & 391 & 781 & 2881 \\\hline
\end{tabular}
}
\end{table}

\section{Hierarchical Imbalanced data learning with Weighted sampling and Label smoothing}
In fact, deep learning-based galaxy image classification tasks are often plagued by the following three typical data characteristics: some classes have a greater degree of similarities with neighboring classes than others, the imbalance in the amount of data between classes, and the discrepancy between the discrete representation of Galaxy classes and the essentially gradual changing of morphology. Therefore, in this section we propose a novel learning method ``Hierarchical Imbalanced data learning with Weighted sampling and Label smoothing (HIWL)". The method is designed based on three parts: hierarchical learning using an efficient model, weighted sampling, and label smoothing. In these three parts, we analyze three typical problems, and combine the dataset Galaxy Zoo-The Galaxy Challenge to give solutions from general to specific. In Section \ref{ssec:Hierarchical learning using efficient models}, we combine the deep network model EfficientNet with the idea of hierarchical learning to learn the features of galaxy images. These images have the characteristic that some classes have a greater degree of similarities with neighboring classes than others. In Section \ref{ssec:weighted sampling}, we use weighted sampling to reduce the negative impact of the imbalanced characteristics of the data. And in Section \ref{ssec:label smoothing}, we use label smoothing to alleviate the machine learning problems caused by the discrepancy between the discrete representation and morphology gradual changing.

\subsection {Hierarchical learning using efficient models}
\label{ssec:Hierarchical learning using efficient models}
In the multi-class galaxy image recognition task, there are few similarities between most classes, and they can be easily distinguished from each other. Only a few classes are difficult to be distinguished from each other and there exist high similarities between them. This kind of phenomenon is referred to as similarity imbalance. The feature extraction capability of the model is particularly important. If the backbone as the foundation is not sufficient to learn high-quality features, the recognition effect of the model would not be satisfactory. Especially, the samples with few similarities between each other appear much more frequently and attract unapproximately much attention from the learning model. And the result is that the model has insufficient ability to identify the classes with high similarities between each other. Thus, feature extraction capability and similarity imbalance brings a challenge to the classification performance of the model. To deal with this kind challenges, we chose the efficient EfficientNet series of models as the backbone in Section \ref{sssec:deep network model EfficientNet} and combine it with the strategy of hierarchical learning in Section \ref{sssec:hierarchical learning} to train the model.

\subsubsection{Deep learning model EfficientNet}
\label{sssec:deep network model EfficientNet}
In this part, we introduce the overall architecture and ideas of the deep network model EfficientNet, give the reflections we made in selecting this series of models, and the specific structure of the EfficientNet-B1 model we selected.

EfficientNet \citep{tan2019efficientnet} is the deep learning model proposed by Google Brain in 2019. Before EfficientNet was proposed, most of the exploration works to improve the performance of neural networks focused on the influence of one of the factors such as network width, depth, and resolution. However, it is shown that the relationship between the three is inseparable, and model performance quickly saturates in this way. The idea of EfficientNet is to use compound expansion to balance the relationship between the above three different factors at the same time, aiming to obtain the optimal model under a certain complexity. It includes a base model B0 obtained by Neural Architecture Search, and extended models B1$\sim$B7 obtained by compound expansion on this basis. Neural Architecture Search \citep[NAS;][]{zoph2016neural} is a subfield of Automatic Machine Learning (AutoML), which aims to achieve more efficient and structured models by designing search spaces, search strategies and performance evaluation strategies. EfficientNet-B0 is obtained from the NAS technique as a benchmark network, which is simple, clean and easy to extend and generalise. EfficientNet-B1 to EfficientNet-B7 are obtained by compound expansion of the benchmark model B0 with different multiplicities for the three factors, corresponding to increasing input resolution (224$\sim$600) and network structure complexity, respectively.

The models with various scales are suitable for the data learning problems with different dataset sizes. For the photometric image dataset Galaxy Zoo-The Galaxy Challenge, we choose B1 from this series of models. It is found that for this dataset, the valid information in most of the sample images is concentrated within a certain range, outside of which there are small bright spots (other galaxies) or black backgrounds. This range is bounded by a central rectangle of 240*240 pixels in the image , and the input resolution for EfficientNet-B1 is 240*240 pixels. In addition, although the size of this dataset is not large, there exists a slight underfitting on the benchmark model B0 due to insufficient network complexity. On the other hand, these are some overfitting on the model B2 or a more complex model. Therefore, this work selected the model B1. 

\begin{figure*}
  \centering
  \includegraphics[width=17.6cm]{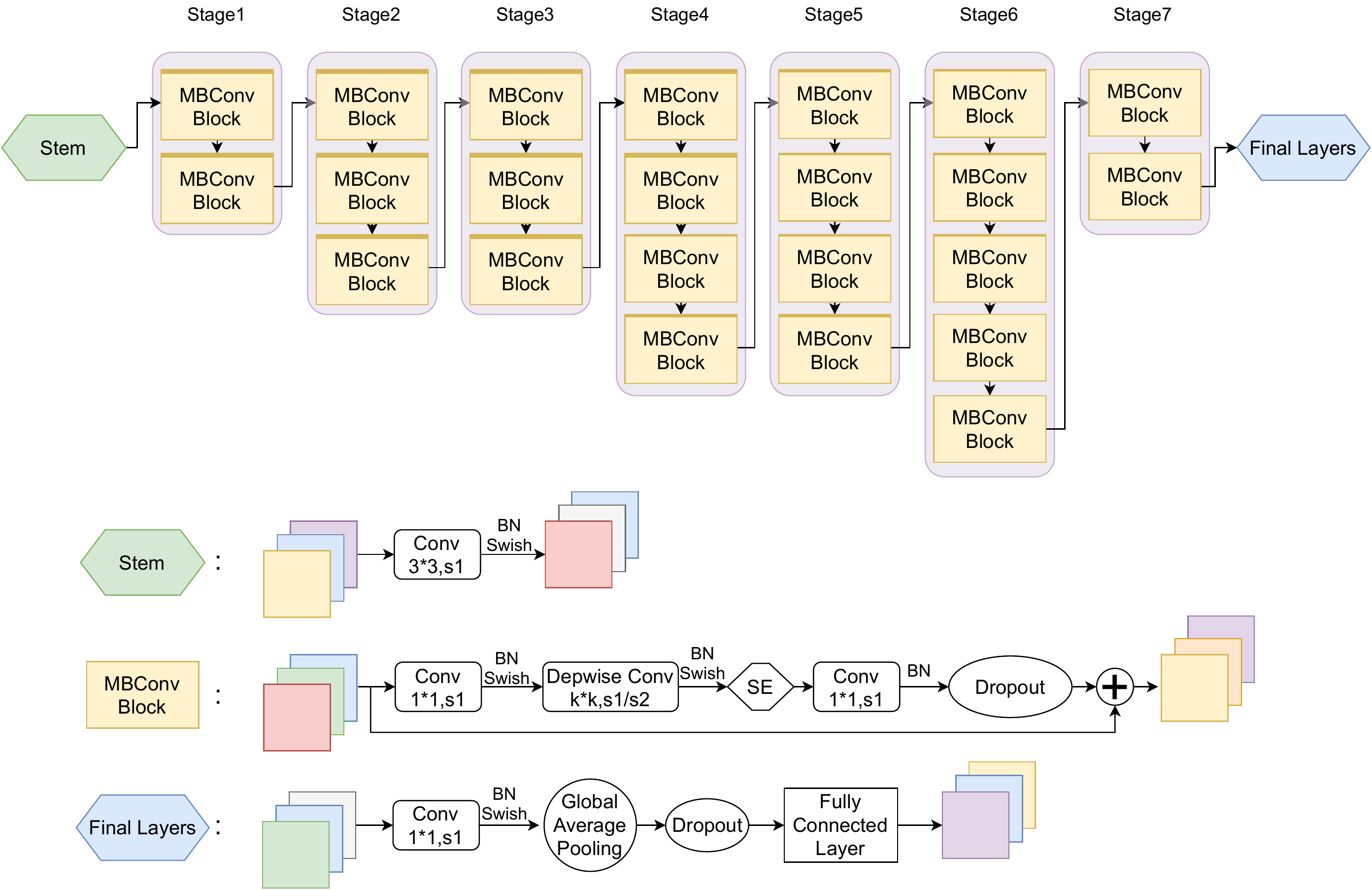}
  \caption{\label{fig:EfficientNet-B1}Structure of EfficientNet-B1. Conv in the figure denotes the convolution operation and s denotes the step size of the convolution. For example, `Conv3*3, s1' denotes a convolution operation with a convolution kernel size of 3*3 and a step size of 1. BN denotes Batch Normalization, which serves to batch normalize the current input and normalize the data under the same scale to speed up model training. Swish refers to the swish activation function, whose expression is y = x * sigmoid(x). }
\end{figure*}

The structure of the EfficientNet-B1 model is shown in Figure \ref{fig:EfficientNet-B1}. This model consists of three modules, namely the Stem module, the MBConv module and the Final Layers module. Following the inference direction, the network first goes through a Stem module. This module serves as a starting point for extracting the initial features and consists of three components, i.e., convolution (kernel size is 3*3), Batch Normalization and swish activation function. It then goes through seven stages, each consisting of a different number of similar modules MBConv stacked to iteratively extract deeper feature information. In EfficientNet-B0$\sim$B7, the stacking numbers of this module are all different. B1 is composed of 23 modules, and there are certain differences in specific parameters in different modules. The DW convolution (Depthwise Convolution) and the channel attention mechanism SE (Squeeze and Excitation) are introduced into the MBConv module, which requires less computation than conventional convolutions. In DW Convolution, a convolution kernel is computed with only one channel's feature map, which greatly reduces the computational effort. However, the connections between channels are not captured by DW convolution, so the SE structure is used to learn the correlation between channels and obtain channel-specific attention. The combination of the two makes the MBConv module lighter and more powerful in integrating channel information, which is the cornerstone of EfficientNet's speed and feature extraction capabilities. The Final layers module, which is the last to pass through, is the end module that consolidates the information and performs the classification. It consists of the following parts: convolution (kernel size is 1*1), Batch Normalization, swish activation functions, global average pooling, dropout layer and fully connected layer.

\subsubsection{Hierarchical learning}
\label{sssec:hierarchical learning}

\begin{figure}
  \centering
  \includegraphics[width=\linewidth]{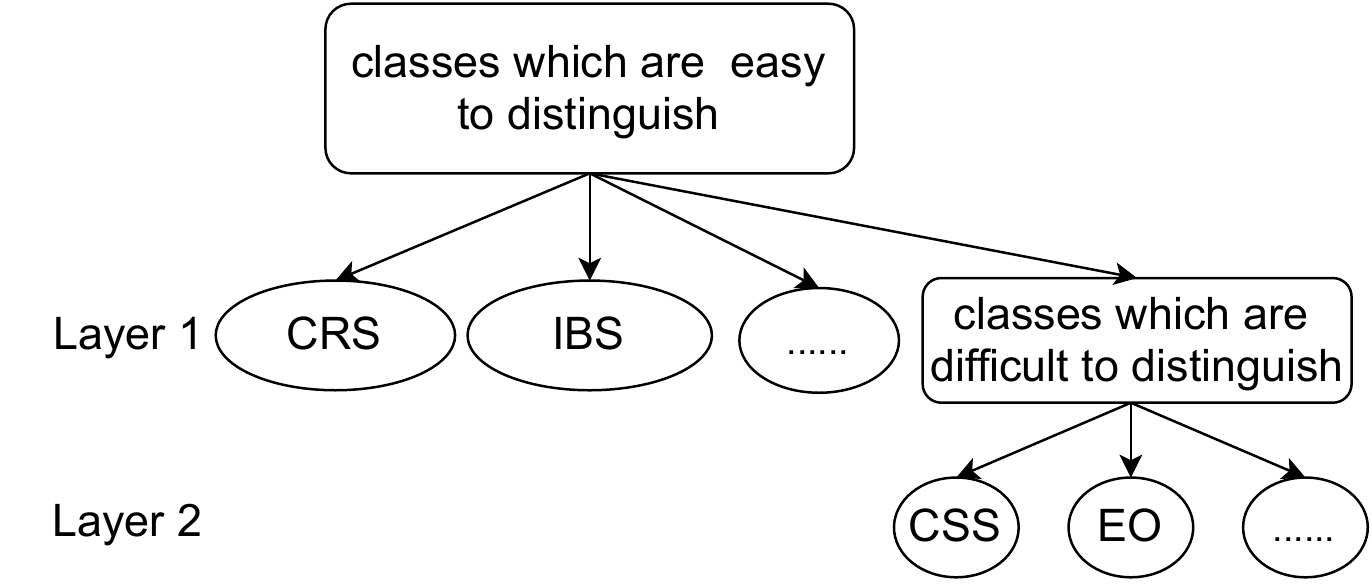}
  \caption{\label{fig:hierarchical learning} Hierarchical learning. The samples from a class which are easy to distinguish share few similarities with the galaxy images from any other class; the samples from a class which are difficult to distinguish share high similarities with the galaxy images from at least one other class. 
 }
\end{figure}

In some image classification tasks, the similarity between two classes of images is different from image pair to image pair. In some image pairs, the samples relatively easily to be misclassified from one class to the other class \citep{song2016two}. Hierarchical learning is an effective method to solve such problems, and the idea is to sequentially classify the data from easy to dificullt. For example, \citet{kim2019machine} used a cascade of 4 two-classification models for hierarchical learning of 5 similar disease classes (the number of layers is 4), which improved the overall disease diagnosis accuracy by 2.8 percent. \citet{ gashi2021hierarchical} used hierarchical learning in a way that first divided the head and face for the five classes of socially active poses and then further divided these two parts, obtaining an improvement of 2$\sim$9 percentage. However, these works used more than 2 classification models, which makes the overall complexity of the model higher. To simplify the problem, this work proposes a novel hierarchical learning scheme with two layers (Figure \ref{fig:hierarchical learning}): in the first layer, all classes which are difficult to distinguish are combined into a separate combinatorial class, which is trained together with other classes which are easy to distinguish; in the second layer, the combinatorial class is split further into some individual classes and the distinction between these classes which are difficult to distinguish is specially learned.

In the Galaxy Zoo-The Galaxy Challenge dataset, particularly, the CSS and EO share the characteristics of being wide in the middle of the image and narrow on their flanks. These characteristics result in some difficulties to discriminate CSS and EO from each other. The shapes of CRS, IBS and SPI are rotundity, flat ellipse, and spiral with arms respectively. Therefore, it is relatively easy to distinguish three objects from each other. 
In the training set, The number of CSS samples is the least (462), the number of EO samples is relatively small (3122), while each of the other three classes has more than 6000 samples. Based on the idea of hierarchical learning, we first treat CSS and EO as a combined class (the number of samples is 3584). This combined class is also easy to be distinguished from the other three classes. Therefore, we classify the samples into four classes in the first layer: the combined class, CRS, IBS, and SPI. This not only makes it easier for the model to learn the difference between the classes, but also alleviates the imbalance between CSS and CRS, IBS, and SPI. The combined classes are then divided further into CSS and EO, with weighted sampling for training in two-classification. Finally, the two models trained above are used sequentially (more details in Section \ref{ssec:Overall flow of the experimental scheme}) to distinguish between the five classes.

\subsection{Weighted sampling}
\label{ssec:weighted sampling}
Data imbalance refers to the large difference in sample size between different classes in a dataset. This characteristic can have a serious negative impact on the overall performance of the CNN \citep{BUDA2018249}. When the classification task uses an imbalanced dataset, the learned model prefers to recognize minority class samples as majority class data. To alleviate the imbalance problem, two popularly used techniques are oversampling and undersampling. Without cooperating with other techniques, experimental investigations show that the oversampling outperforms the downsampling method \citep{mohammed2020machine}. \citet{krawczyk2016evolutionary} proposed an algorithm combining boosting and undersampling, which has high application value in breast cancer tumor diagnosis. The disadvantages of oversampling are the overfitting tendency and the high time cost; undersampling may make the model underfit and under-learning of features due to insufficient amount of data. Therefore, this work proposed a weighted sampling approach to alleviate the imbalance effects better.

The idea of weighted sampling is to apply a weight to each sample data in the imbalanced data, and this weight value is the reciprocal of the number of samples in this class. It means that samples from the minority class get higher weight, and samples from the majority class get lower weight. Samples with high weights have a higher probability of being sampled repeatedly but have fewer number of samples coming from the same class; samples with low weights have a lower probability of being sampled but have a larger number of samples coming from the same class. As a result, the number of samples of each class obtained by weighted sampling is approximately balanced. This idea can be expressed as follows:

\vspace{-0.5cm}
\begin{equation}
w_i=\frac{1}{N_i},    
\end{equation}

\vspace{-0.5cm}
\begin{equation}
p_i=\frac{w_i}{\sum_{j=1}^{k}{w_j\times N_j}}=\frac{w_i}{k},  
\end{equation}

\vspace{-0.5cm}
\begin{equation}
E(C=c_i,M=m)=p_i\times N_i\times m=\frac{m}{k},   
\end{equation}
where $k$ represents the number of class, $N_i$ is the number of samples from class $c_i$, $w_i$ is the weight of each sample in class $c_i$, $p_i$ is the probability that each sample in class $c_i$ is sampled, and $m$ is the number of expected samples. The expectation of the number of samples sampled from any class is $E\left( C = c_{i},M = m \right) = \frac{m}{k}$. Therefore, the computed dataset is balanced.

In the dataset Galaxy Zoo-The Galaxy Challenge, there is an imbalanced relationship between CSS and the other four classes. The imbalance between CSS, EO and CRS, IBS, SPI has been mitigated by the first layer of hierarchical learning. The second layer is devoted to recognizing CSS and EO, and the imbalance problem between them is dealt with the weighted sampling scheme. In the second layer of hierarchical learning, when CSS and EO are fed into the two-classification model, weighted sampling is added due to the imbalance in the amount of data between the two. In the training phase, the weight of each sample of CSS obtained by weighted sampling is 1/462, and the weight of each sample of EO is 1/3122, the corresponding probability of being selected is 1/924 and 1/6244, respectively. It means that each CSS sample is easier to be sampled, but since EO belongs to the majority class(3122) in the CSS-EO classification problem, the sampling probability of any class is 1/2. Therefore, the two classes of samples tend to be in balance. However, this repeatable sampling operation may yield duplicate samples. Therefore, we will add online data augmentation after sampling (described in more detail in Section \ref{ssec:data preprocessing}). Due to the random nature of the data augmentation, there isn't any duplicate samples in the final computation even if one galaxy image is sampled more than twice. Therefore, this strategy ensures the diversity of the final data. 

\subsection{Label smoothing}
\label{ssec:label smoothing}
In traditional supervised machine learning, labels are often encoded in a one-hot fashion (in Equation \ref{eq:one-hot}).

\vspace{-0.5cm}
\begin{equation}
\label{eq:one-hot}
{\hat{y}}_{i} = \left\{ \begin{matrix}
{1,~~i = target} \\
{0,~~i \neq target} \\
\end{matrix} \right.
\end{equation}
Based on this one-hot label, the target class predictions will converge to 1 and the non-target predictions converge to 0 in minimizing the loss . This label encoding ignores the relationship between the real class and other classes, and it cannot guarantee the generalization ability of the model, which sometimes makes the model prone to overfitting \citep{Zhang2019BagOF}. In addition, this drawback is amplified when encountering cases with wrong labels. 

When using one-hot encoding as a training label for multi-class galaxy morphology recognition, the labels are only discrete representations like 0 and 1. However, galaxy morphology is essentially gradual changing, which means that the labels should not be just this either/or configuration. Specifically, for the dataset Galaxy Zoo-The Galaxy Challenge, in the Hubble classification standard, there is a gradual changing characteristic between CRS and SPI. And in the classification standard of Galaxy Zoo 2, CRS, CSS, and IBS belong to smooth galaxies, and there are some gradual changing morphological characteristics on them. Based on one-hot encoding, the model only learns the typical features of each class, but not the gradual changing characteristics between the classes. This situation results that the model has a higher error rate when recognizing transitional morphological type samples.

Label smoothing, a regularization method in the field of machine learning, was originally proposed by \citet{szegedy2016rethinking}. The method has been applied to several fields, such as image classification \citep{hou2019multi}, image segmentation \citep{islam2021spatially}, machine translation \citep{liang2022implicit}, and speech recognition \citep{zheng2020homophone}. This is done by making an improvement to one-hot labels (in Equation \ref{eq:label smothing}) by introducing a smaller parameter $\alpha$ and $K$ as the number of predicted classes, resulting the labels for target class locations are $1-\alpha$ and the labels for non-target class locations are $\alpha/(K - 1)$. \citet{ Mller2019WhenDL} explains from a representational visualization perspective that label smoothing decreases the intra-class distance of samples in space and increases the inter-class distance of samples in space, which facilitates the calibration of the model.

\vspace{-0.5cm}
 \begin{equation}
 \label{eq:label smothing}
{\hat{y}}_{i} = \left\{ \begin{matrix}
{1 - \alpha,~~i = target} \\
{\alpha/(K - 1),~~i \neq target} \\
\end{matrix} \right.
\end{equation}

\vspace{-0.5cm}
\begin{equation}
 \label{eq:softmax}
p_{c=i}=\frac{exp(z_{c=i})}{\sum_{j=1}^{k}{exp(z_{c=j})}}\\
\end{equation}

When using softmax for multi-classification tasks, the predicted value of the model on the target class $i$ is described as Equation \ref{eq:softmax}. Where $c$ represents the class, $p$ is the predicted value of a certain class, $z$ is the input logical vector of $softmax$, and $k$ represents the dimension of the logical vector. During the training of the model under the one-hot encoding method, the predicted value of the target class i $p_{c=i}$ keep tending to 1, and the predicted value of the non-target class $p_{c \neq i}$ keep tending to 0. It means that $z_{c=i}$ tends to positive infinity, and $z_{c \neq i}$ tends to negative infinity. At this time, since the input $x$ is a fixed value, the value of the weight $w$ or the bias $b$ will be correspondingly huge, which is easy to cause overfitting. After using label smoothing, the goal of $p_{c=i}$ is to tend to a fixed value of less than 1. 
This makes $z_{c=i}$, $z_{c \neq i}$ no longer tend to infinity, and w, b are no longer optimized when reaching a certain value. Therefore, label smoothing suppresses overfitting and performs better than one-hot. In addition, label smoothing makes each bit of the label code has a value of $0-\alpha$ to reflect the gradual changing characteristic between the classes. For example, the original one-hot encoding used by CSS is (0, 0, 1, 0, 0), and after smoothing the label with the parameter $\alpha$=0.1, the encoding becomes (0.025, 0.025, 0.9, 0.025, 0.025 ). The 0.025 in the code represents the relationship between this class and other classes to a certain extent, and it also means that the sample has the possibility of belonging to each class. Therefore, using label smoothing makes the label more suitable for gradual changing in a sense, and the learned model will be more reasonable.

\section{Experiment}
\label{sec:experiment}
To explore the effectiveness of the proposed method HIWL, in this section we conducted an experimental study. We applied HIWL to a real galaxy dataset, to verify the classification effectiveness and compare the model with other galaxy classification models. Section \ref{ssec:Overall flow of the experimental scheme} describes the analysis and experimental procedure for applying HIWL to the specific galaxy dataset, Galaxy Zoo-The Galaxy Challenge. As an indispensable part of the experiment, the data augmentation is described in detail in Section \ref{ssec:data preprocessing}. Finally, Section \ref{ssec:training strategies and experimental parameters} presents some details of the experiments, such as parameter settings and specific training strategies.

\subsection{Overall design of the experiment}
\label{ssec:Overall flow of the experimental scheme}
Before the experiments, we analyzed the dataset we used and learned that this dataset has the following characteristics: (1) CSS are very similar to EO in morphology, while other galaxies are highly differentiated from each other; (2) the amount of data for CSS is much less than the amount of data for the other four classes; (3) the galaxy images in the dataset have gradual changing characteristics in morphology; and the data may be mislabeled, because the amount of source data is huge and the labeling is done after short-term learning by volunteers. We applied our learning method HIWL to the experiments on this dataset. That is, in the overall design of the experiment, hierarchical learning is used to learn the difference between similar classes and alleviate the data imbalance effects, weighted sampling is used to deal with the imbalance problem, and label smoothing is used to learn the gradual changing characteristics between classes.

\begin{figure*}
  \centering
  \includegraphics[width=17.2cm]{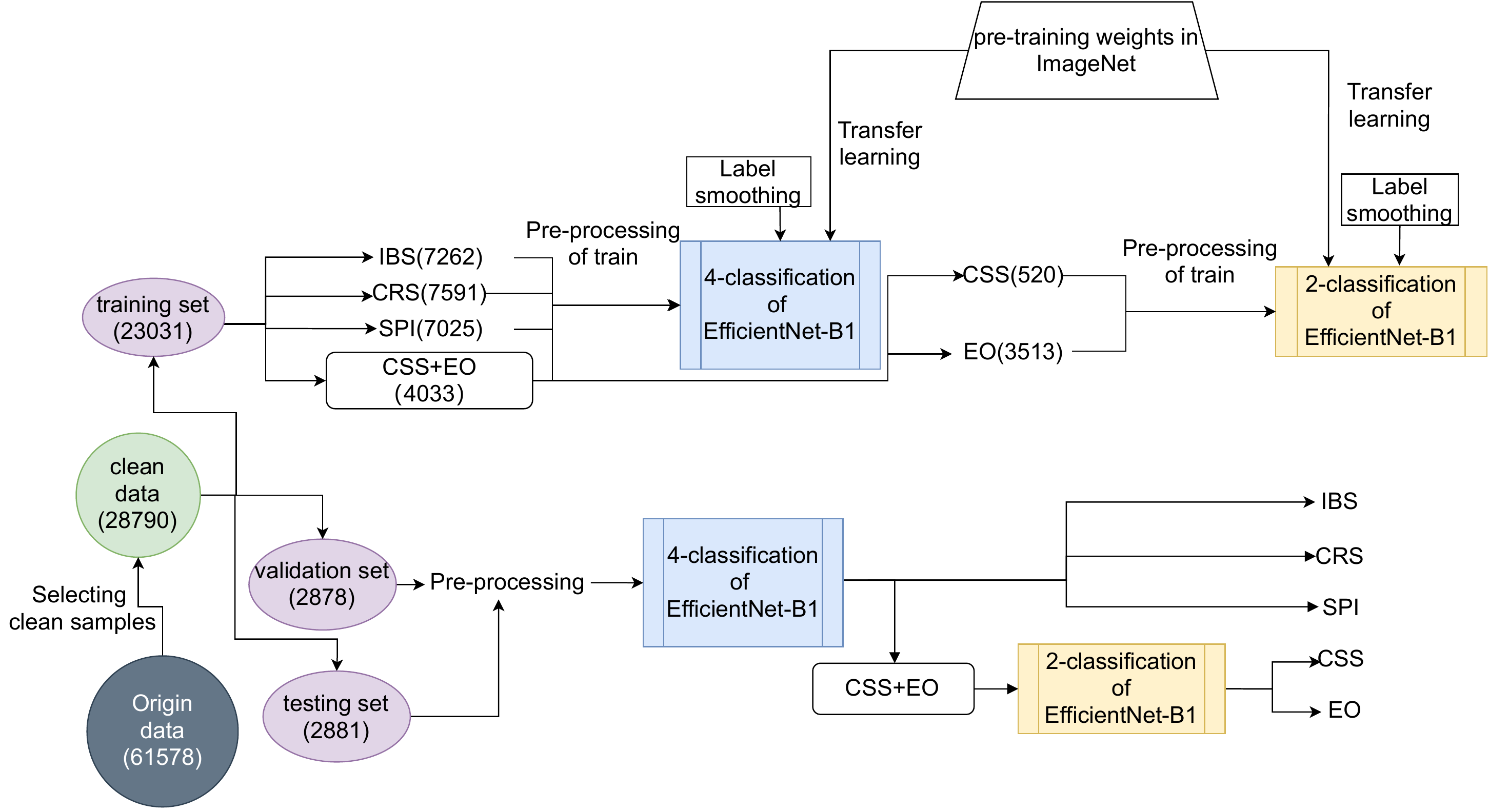}
  \caption{\label{fig:experiment}The overall design of the experiment. The design as a whole incorporates the following four operations: (1) dataset selection and division; (2) hierarchical learning; (3) data preprocessing; (4) learning strategies in the training phase.}
\end{figure*}

The overall design of the experiment is shown in Figure \ref{fig:experiment}. First, we filter the original samples (61578 galaxy images) according to the clean sample selection rules to obtain clean sample data (28790 images), and divide them into training set, validation set and test set. In the training phase, we adopt the idea of hierarchical learning to classify the training data into four classes, namely CRS, IBS, SPI and a combined class of CSS and EO. The samples of these 4 classes will be processed using various data preprocessing procedures (details in Section \ref{ssec:data preprocessing}), and two training strategies (details in Section \ref{ssec:training strategies and experimental parameters}). After that, it is sent to the four-classification model with EfficientNet-B1 as the main body for training. Furthermore, for the two classes that are difficult to be distinguished from each other, namely CSS and EO, we train a two-classification model in the same way as above. In the validation phase or test phase, we put the samples into the trained four-classification model after pre-processing. And if the sample is judged to be one of the non-combination classes, the result is the final prediction result; if the sample is judged to be a combination class, we need to pass it through the trained two-classification model to get the final prediction result.
 
\subsection{Data augmentation}
\label{ssec:data preprocessing}
\begin{figure}
  \centering
  \includegraphics[width=\linewidth]{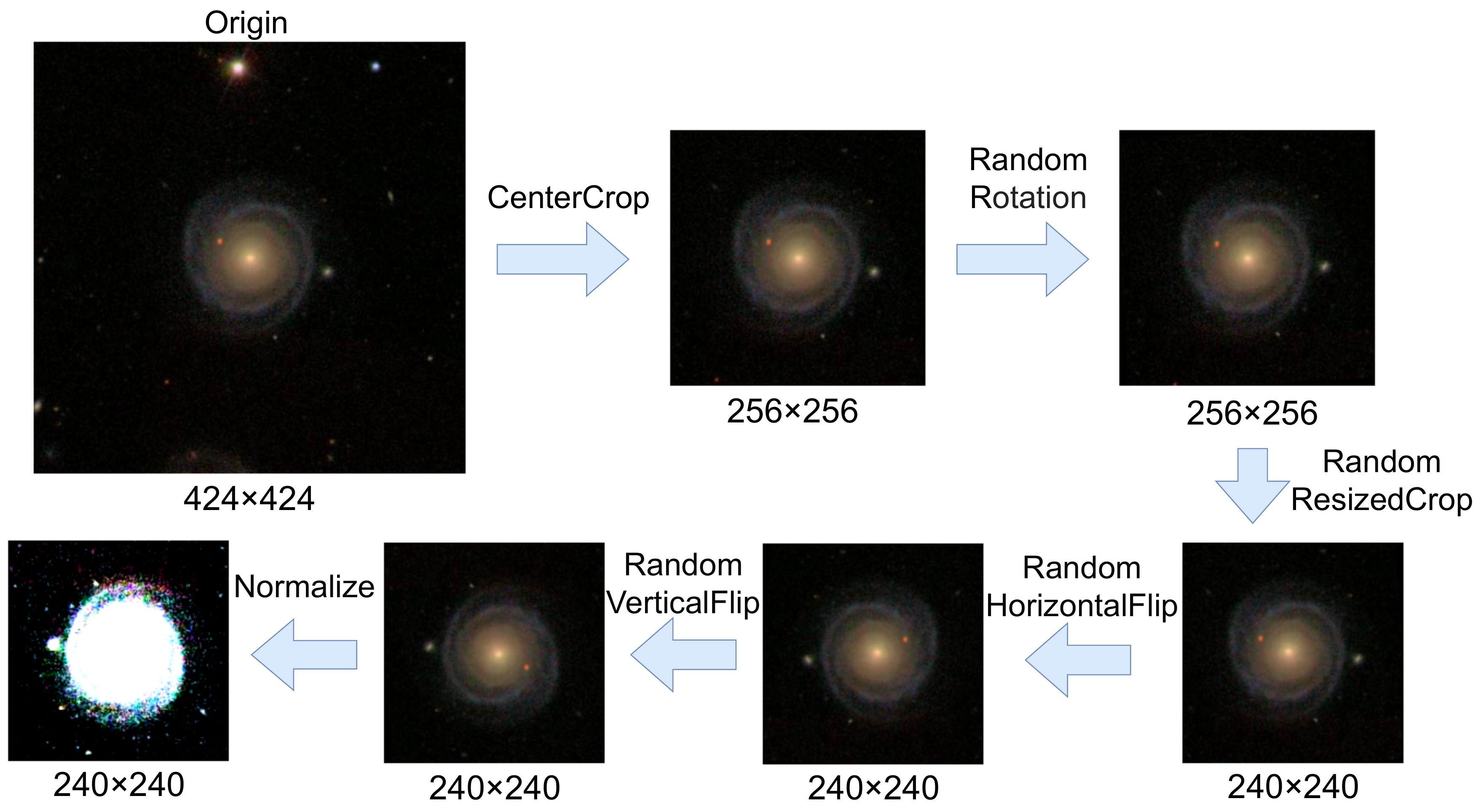}
  \caption{\label{fig:preproceding} Data augmentation of training phase. The original image of the sample is used as the final input of the network structure after the following operations: center cropping to 256$\times$256, random rotation, random scaling and cropping to 240$\times$240, random horizontal flip, random vertical flip, and finally normalization.}
\end{figure}

For deep learning, a total of 23031 training samples is clearly not enough to train a sufficiently robust model. Therefore, it is necessary to preprocess the training data using some data augmentation techniques. Considering the diversity of galaxy morphology and the effective information concentration range of most of the images in this dataset (within 240$\times$240 pixels in the centre), we enrich the sample data through some data augmentation operations (as shown in the Figure \ref{fig:preproceding}). These data augmentation operations include center cropping and random operations such as rotation, scaling, horizontal flipping, and vertical flipping. 

Suppose the original training set is expressed as $S_{tr}=\{ (\mathbf{x}_i,y_i), i=1, 2, \cdots, 23031\}$ (Section \ref{ssec:dataset selection}). Where 23031 denotes the number of training samples, $\mathbf{x}_i$ denotes the $i$th galaxy image, and $y_i$ denotes the class of $\mathbf{x}_i$. The method of model learning in this paper is an iterative optimization scheme. Each complete iteration of the training set is called an epoch, and the model parameters are updated in each epoch based on an approximate dataset of the training sample set $S_{tr}$. Assume that the model has experienced a total of $N_ {e}$ epochs of iterative learning, then the learning process of the four-classification model in this paper is summarized as follows:

\begin{itemize}
    \item For $n$ from 1 to $N_{e}$ 
    \item Generate an approximate dataset of the sample set $S_{tr}$ using the data augmentation method (Figure \ref{fig:preproceding}). $S_{tr}^n=\{ (\mathbf{z}_i,y_i), i=1, 2, \cdots, 23031\}$. The number of samples in $S_{tr}^n$ is the same as the number of samples in the training set $S_{tr}$, and the class of both sample $\mathbf{z}_i$ and sample $\mathbf{x}_i$ is $y_i$. The sample $\mathbf{z}_i$ is generated from the sample $\mathbf{x}_i$ by the data augmentation method of Figure \ref{fig:preproceding}: (1) an image $\mathbf{v}_{i}^{(1)}$ is generated from the sample $\mathbf{x}_i$ by CenterCrop; (2) do a RandomRotation on image $\mathbf{v}_{i}^{(1)}$ to generate an image $\mathbf{v}_{i}^{(2)}$; (3) do a RandomResizedCrop on image $\mathbf{v}_{i}^{(2)}$ to generate an image $\mathbf{v}_{i}^{(3)}$; (4) do a RandomHorizontalFlip on image $\mathbf{v}_{i}^{(3)}$ to generate image $\mathbf{v}_{i}^{(4)}$; (5) do a Random VerticalFlip on image $\mathbf{v}_{i}^{(4)}$ to generate image $\mathbf{v}_{i}^{(5)}$; (6) do a normalization on image $\mathbf{v}_{i}^{(5)}$ to generate the image $\mathbf{z}_{i}$.
    \item use the dataset $S_{tr}^n$ to learn the model.
    \item Perform the next round of iterations.
\end{itemize}
Where the occurrence probability of the RandomHorizontalFlip in step (4) and the RandomVerticalFlip in step (5) of the data augmentation are both 0.5, and the occurrence probability of each operation in the other steps is 1.

In the above learning process, $S_{tr}^{n}$ is generated independently in each iteration by the data augmentation method. This data augmentation method is called the online data augmentation method. In contrast, the offline data augmentation method generates a certain augmented dataset $S_{tr}^{'}$ from the training set $S_{tr}$ prior to iterative training, and uses $S_{tr}^{'}$ for computation in each iteration of learning. In total, $N_e\times 23031 = 2832813$ samples were generated by the online data augmentation method throughout the iterative learning process. This greatly enhances the diversity of the training data and ensures the generalization ability of the learning results. At the same time, only 23031 samples are used in each training round, which is the same number of samples as the original training set $S_{tr}$, ensuring a manageable computational burden. In addition, considering the data capacity and computational burden issues, we are not able to generate 2832813 augmented samples and use them for model training in the offline data augmentation scheme.

In the above four-classification model training, each augmented dataset $S_{tr}^n$ was generated from the original training set $S_{tr}$ using sampling without replacement: each sample in $S_{tr}$ produces a unique augmented sample to put into $S_{tr}^n$. However, the typical characteristics of two-classification model learning on CSS and CO is the imbalance of the data, with a data ratio of 462:3122$\approx$1:6.76 (Table \ref{table:data distribution}). Therefore, sampling with replacement is adopted when generating the augmented dataset $S_{tr, CSS-EO}^n=\{ (\mathbf{z}_i,y_i), i=1, 2, \cdots, 3584\}$ from the origin training set $S_{tr, CSS-EO}=\{ (\mathbf{x}_j,y_j), j=1, 2, \cdots, 3584\}$: for any $i=1, 2, \cdots, 3584$, a sample $\mathbf{x}$ is randomly selected from $S_{tr, CSS-EO}$ using the weighted resampling method of Section \ref{ssec:weighted sampling}, assuming its class is $y$; then using steps (1)-(6) of the previous data augmentation by $ \mathbf{x}$ to generate the image $\mathbf{z}$, such that $\mathbf{z}_i = \mathbf{z}$, $y_i= y$; and $(\mathbf{z}_i, y_i)$ as the $i$th sample of $S_{tr,CSS-EO}^n$ to form the augmented dataset. This is still an online data augmentation method. According to the exploration in Section \ref{ssec:weighted sampling}, the augmented sample set $S_{tr,CSS-CO}^n$ maintains a balance between CSS and EO.

\subsection{Training strategies and experimental parameters}
\label{ssec:training strategies and experimental parameters}
\subsubsection{Training strategies}
\label{sssec: training strategy}
This section described the schemes for model initialization and optimal model selection. HIWL includes two sub-models, namely, a four-classification model and a two-classification model. In the initialization phases of both sub-models, transfer learning is adopted. The approach is that during initialization, the model loads the pre-training weights that PyTorch officially trained in the Imagenet dataset. In this way, the model can converge earlier, and has a strong feature extraction ability at the beginning.  

For optimal model selection, the models with top three accuracies will be retained for obtaining the final model. A complete iteration means that all samples in a set are learned by the model once. Each complete iteration of the training set is called an epoch, and each classifier needs to be learned in a number of epochs. Suppose $N_e$ denotes the total number of epochs of a model, $N_e^4$ denotes the total number of epochs of the four-classification model, and $N_e^2$ denotes the total number of epochs of the two-classification model. After each epoch, a classifier and its classification accuracy on the validation set can be computed. Thus, after $N_e$ rounds of iterations, $N_e$ candidate classifiers and their validation accuracy are obtained. An overly large $N_e$ can increase the risk of overfitting, and an inappropriately small $N_e$ usually results in underfitting. These two phenomena have a negative impact on validation accuracy. Therefore, a large epochs number (such as 1000) would be set in advance as an upper bound for $N_e$ and the training procedure will be terminated when the loss value on the validation set no longer decreases and the validation accuracy no longer increases. At this point, the number of passed epochs is the $N_e$. Based on this scheme, the $N_e^4$ and $N_e^2$ are chosen as 123 and 96 respectively. The proposed galaxy image classification system HIWL consists of two subclassifiers: one four-classification model, and one two-classification model. In establishing the HIWL, the combinational effects between the subclassifiers should be considered. Therefore, when choosing the best model, the models with the top three validation accuracies in 123 candidate four-classification models will be retained, and the models with the top three validation accuracies in 96 candidate two-classification models will be retained. During the 123 epochs, the top three validation accuracies of four-classification models are 97.84\% (at the 90th epoch), 98.02\% (at the 98th epoch), 97.91\% (at the 120th epoch). During the 96 epochs, the top three validation accuracies of two-classification models are 95.30\% (at the 47th epoch), 95.08\% (at the 57th epoch), 95.08\% (at the 62nd epoch). As a result, nine HIWL models are generated, and the HIWL model with the highest validation accuracy is the final model.

\subsubsection{Experimental parameter settings}
\begin{figure}

  \subfigure[Training loss]{\includegraphics[width=\linewidth]{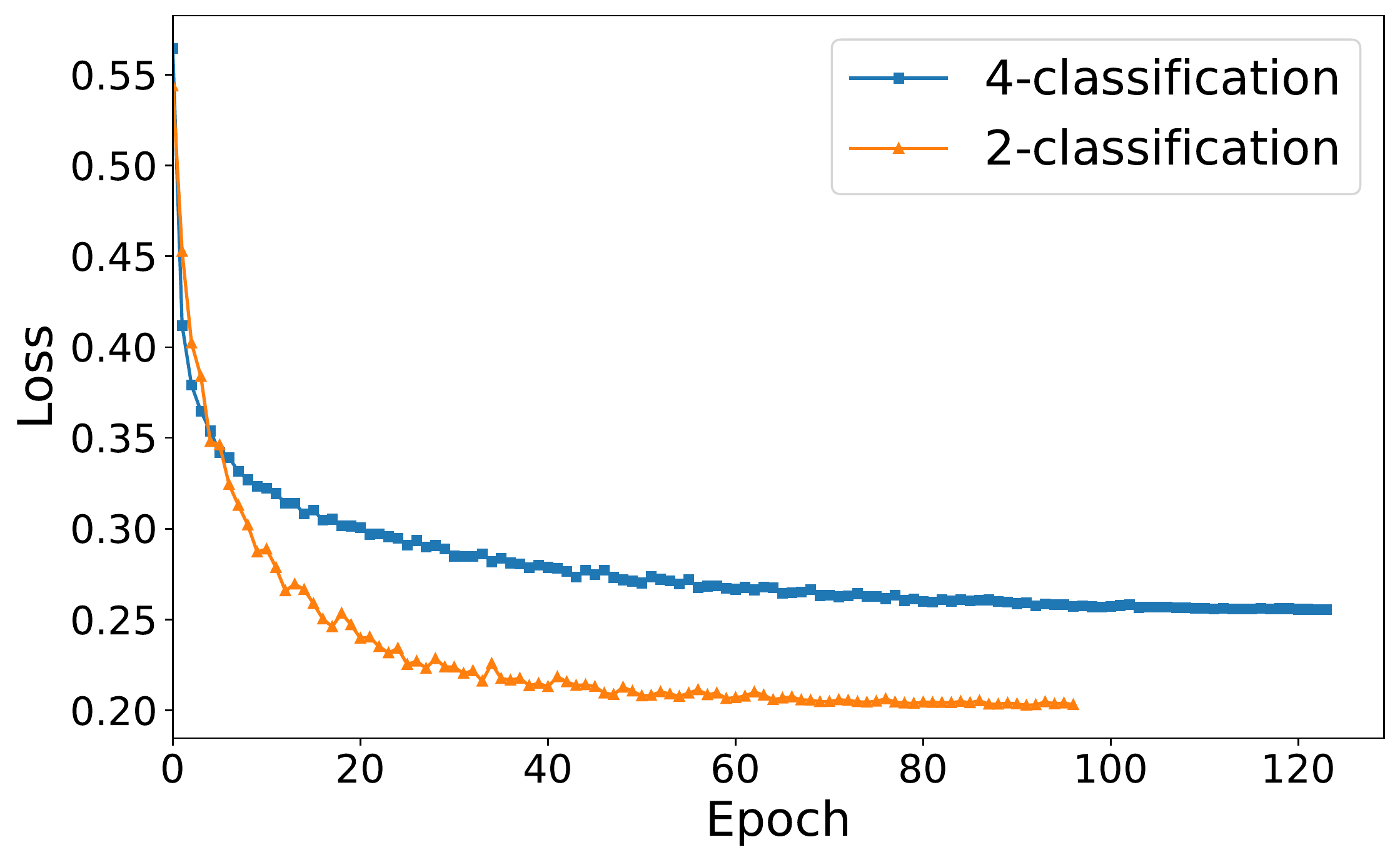}\label{loss}}
  \subfigure[Validation accuracy]{\includegraphics[width=\linewidth]{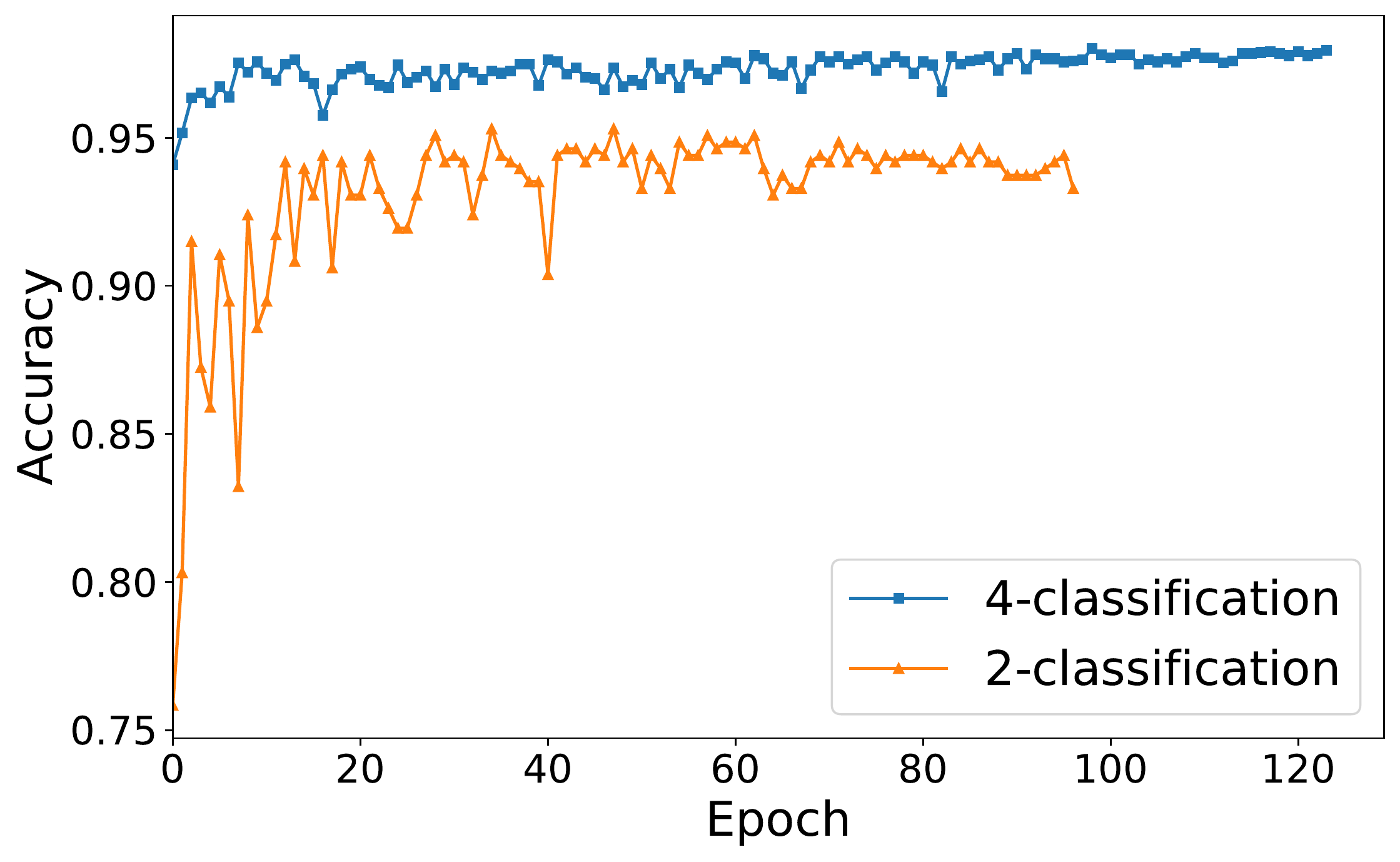}\label{acc}}
  \caption{\label{fig:loss and acc}Training loss and validation accuracy curves for two training phases. In subgraph (a) or (b), the horizontal represents the epoch, and the vertical represents the loss or accuracy. The total number of epochs for the 4-classification model is 123, and the total number of epochs for the 2-classfication model is 96.}
\end{figure}

Different parameter settings often lead to different experimental results, and our specific experimental parameters are as follows. The deep learning framework we use is PyTorch, the GPU is Nvidia RTX 2060 and the memory is 6GB. For the four-classification model, we iterated 123 epochs based on transfer learning, the learning rate is 0.005, the learning rate decay strategy of cosineAnnealing is adopted, the batch size is 24, and the label smoothing parameter $\alpha$=0.05. For the two-classification model, we iterated 96 epochs based on transfer learning, the learning rate is 0.005, the learning rate decay strategy of cosineAnnealing is adopted, the batch data size is 24, and the label smoothing parameter $\alpha$=0.05. Another important problem is the determination of an appropriate number of epochs for training the model (the time to stop the training procedure). To do this, the relationship between training loss and the number of epochs and the relationship between validation accuracy and the number of epochs are investigated (Figure \ref{fig:loss and acc}). In Figure \ref{loss}, the loss values of the sub-models both decrease and then converge to low values when the training process does not reach the stopping point. During this period, the corresponding validation accuracies of the sub-models (Figure \ref{acc}) both increase and then stabilise at high values, without any significant decrease. Therefore, both sub-models are not overfitting during the training phase. And the sub-models have a higher degree of data fit and accuracies where both loss values and validation accuracies are stable. Furthermore, when selecting the optimal model (details in Section \ref{sssec: training strategy}), the following candidate sub-models with high validation accuracy are all retained where the losses are at low values. During the 123 epochs, the four-classification models with top three validation accuracies are retained at the 90th epoch (97.84\%), the 98th epoch (98.02\%), and the 120th epoch (97.91\%). During the 96 epochs, the two-classification models with top three validation accuracies are retained at the 47th epoch (95.30\%), the 57th epoch (95.08\%), and the 62nd epoch (95.08\%). Therefore, the choice of epochs that 123 for four-classification and 96 for two-classification of the model is appropriate.

\section{Results and discussion}
\subsection{Evaluation metrics}
The model evaluation metrics used in this experiment are Accuracy (in Equation \ref{eq:accracy}), Recall (in Equation \ref{eq:recall}), Precision (in Equation \ref{eq:precision}) and F1-Score (in Equation \ref{eq:f1}). Accuracy means the ratio of the number of correctly predicted samples to the total number of samples, and is a statistic for all samples. Specifically in galaxy image classification, it refers to the number of galaxy images correctly classified by the model as a ratio of the number of all galaxy images.

\vspace{-0.5cm}
\begin{equation}
\label{eq:accracy}
Accuracy=\frac{N_{TP}+N_{TN}}{N_{TP}+N_{TN}+N_{FP}+N_{FN}}
\end{equation}
where $N_{TP}$ is the number of true positives, $N_{TN}$ is the number of true negatives, $N_{FP}$ is the number of false positives and $N_{FN}$ is the number of false negatives.

Recall means the ratio of the number of correctly predicted samples in a class to the total number of such samples, and is a statistic for all samples in a class. Specifically in galaxy image classification, it refers to the ratio of the number of galaxy images in a class (e.g., CRS) that are correctly classified by the model to the number of all galaxy images in that class.

\vspace{-0.5cm}
\begin{equation}
\label{eq:recall}
Recall=\frac{N_{TP}}{N_{TP}+N_{FN}}
\end{equation}

Precision means the ratio of the number of correctly predicted samples in a class to the total number of samples predicted as such, and is a statistic for all samples predicted as a certain class. Specifically in classification of galaxy images, it refers to the ratio of the number of galaxy images of a certain class (e.g., CSS) that are correctly classified by the model to the total number of galaxy images classified by the model into this class.

\vspace{-0.5cm}
\begin{equation}
\label{eq:precision}
Precision=\frac{N_{TP}}{N_{TP}+N_{FP}}
\end{equation}

Therefore, recall and precision have different emphases. Furthermore, the F1-Score is based on the harmonic mean of precision and recall, which is an evaluation metric after weighing the two.

\vspace{-0.5cm}
\begin{equation}
\label{eq:f1}
F1\ Score=2\ast\frac{Precision\ast Reacall}{Precision+Reacall}
\end{equation}

Accuracy measures the prediction of a global sample and gives a general view of the model's performance, but it does not provide a more detailed local view of the model's prediction for a certain class. Therefore, recall, precision and F1-Score need to be considered. When the cost of False Negative (FN) is high, i.e., when misclassification of samples in a class into other classes would have serious consequences , the focus should be on improving the recall of that class. When the cost of False Positive (FP) is high, i.e., when being misclassified into a certain class has serious consequences, the focus should be on improving the precision of that class. The F1-Score is a combination of recall and precision and is not biased towards either, which can be applied to more general situations.

\subsection{Results presentation of HIWL}
We save the models with top three accuracies for the four-classification and two-classification models during training. Then, one of the two types of models is selected, and after combined validation, the model HIWL with the highest combined accuracy is obtained. The accuracy of this model HIWL on the test set is 96.32\%. The corresponding confusion matrix is shown in Table \ref{table:confusion matrix}, and Table \ref{table:evaluation} denotes its various evaluation metrics (recall, precision, F1-Score) on the test set.

\begin{table}
\centering
\caption{\label{table:confusion matrix}Confusion matrix of HIWL on the test set. The vertical represents the true class and the horizontal represents the predicted class of this model.}
\begin{tabular}{llllll}
\hline
 & CRS & IBS & CSS & EO & SPI \\ \hline
CRS & 813 & 26 & 0 & 0 & 5 \\
IBS & 25 & 777 & 0 & 0 & 5 \\
CSS & 0 & 6 & 41 & 11 & 0 \\
EO & 0 & 1 & 11 & 376 & 3 \\
SPI & 2 & 8 & 0 & 3 & 768 \\ \hline
\end{tabular}
\end{table}

\begin{table}
\centering
\caption{\label{table:evaluation}Recall, precision and F1-Score of HIWL on the test set. Avg represents the arithmetic mean of the corresponding metrics for all classes.}
\begin{tabular}{cccc}
\hline
Class & Recall & Precision & F1-Score \\ \hline
CRS & 0.9633 & 0.9679 & 0.9656 \\
IBS & 0.9628 & 0.9499 & 0.9563 \\
CSS & 0.7069 & 0.7855 & 0.7455 \\
EO & 0.9616 & 0.9641 & 0.9629 \\
EPI & 0.9834 & 0.9834 & 0.9834 \\
Avg & 0.9156 & 0.9302 & 0.9227 \\ \hline
\end{tabular}
\end{table}

From Table \ref{table:confusion matrix}, it is shown that the ratio of misclassification is less than 5\% for all classes except the CSS. The relatively large ratio of misclassifications for CSS is due to the fact that CSS and EO are too similar and the original sample data for CSS is too small. It is shown from the Table \ref{table:evaluation} that except for the CSS, the recall and F1-Score of other 
classes are over 0.956, and the precision is over 0.949. In particular, the precision of CSS is close to 0.8, and the recall and F1-Score are also over 0.7.

In theory, the hierarchical learning mechanism will increase the complexity of a galaxy classification model. To explore whether this approach is advisable, we compared the HIWL with a simpler model. This simpler model consists of a single layer with weighted sampling and label smoothing (SLWSLM), and is designed by removing the hierarchical learning mechanism from HIWL. The experimental results on the test set are presented in Table \ref{table:SLWSLM}. It is shown that the hierarchical learning mechanism has some evident impacts on galaxy image recognition. In particular, the improvement is 0.42\% for the overall accuracy, and 8.62\% and 2.81\% on the recall improvements for minority classes CSS and EO. The data used in training the second-layer sub-model belongs to two minority classes and accounts for a small subset of the entire dataset. Therefore, the increased training time is short(48 seconds/epoch). Specifically, the training time of each epoch of HIWL and SLWSLM is 369 seconds and 321 seconds, respectively. Furthermore, the EfficientNet has low complexity and fast convergence speed than general backbone models (such as resnet). Therefore, the hierarchical learning scheme is desirable.

\begin{table}
\caption{\label{table:SLWSLM}The desirability of using a hierarchical classification mechanism. This evaluation on test set is conducted by comparing the HIWL with a simpler model SLWSLM (a single layer (model) with weighted sampling and label smoothing). The SLWSLM is designed by removing the hierarchical learning mechanism from HIWL.}
\begin{tabular}{ccccc}
\hline
\multirow{2}{*}{} & \multicolumn{2}{c}{HIWL} & \multicolumn{2}{c}{SLWSLM} \\ \cline{2-5} 
 & Recall & Overall acc & Recall & Overall acc \\ \hline
CRS & 0.9633 & \multirow{5}{*}{96.32\%} & 0.9763 & \multirow{5}{*}{95.90\%} \\
IBS & 0.9628 &  & 0.9579  &  \\
CSS & 0.7069 &  & 0.6207  &  \\
EO & 0.9616 &  & 0.9335  &  \\
SPI & 0.9834 &  & 0.9795  &  \\ \hline
\end{tabular}
\end{table}

To explore the effect of label smoothing, we compared the HIWL with an another simpler model. This simpler model consists of a hierarchical model with weighted sampling (HMWS), and is designed by removing the label smoothing mechanism from HIWL. The experimental results on test set are presented in Table \ref{table:HMWS}. It is shown that the label smoothing mechanism has positive impacts on overall accuracy and recall of galaxy image recognition. In particular, the improvement is 0.28\% for the overall accuracy, and 0.12\%, 3.45\%, 1.79\%, 0.13\% on the recall improvements for IBS, CSS, EO, SPI, respectively.

\begin{table}
\caption{\label{table:HMWS}The desirability of using a label smoothing mechanism. This evaluation on the test set is conducted by comparing the HIWL with a simpler model HMWS (a hierarchical model with weighted sampling). The HMWS is designed by removing the label smoothing mechanism from HIWL.}
\begin{tabular}{ccccc}
\hline
\multirow{2}{*}{} & \multicolumn{2}{c}{HIWL} & \multicolumn{2}{c}{HMWS} \\ \cline{2-5} 
 & Recall & Overall acc & Recall & Overall acc \\ \hline
CRS & 0.9633 & \multirow{5}{*}{96.32\%} & 0.9668 & \multirow{5}{*}{96.04\%} \\
IBS & 0.9628 &  & 0.9616  &  \\
CSS & 0.7069 &  & 0.6724  &  \\
EO & 0.9616 &  & 0.9437  &  \\
SPI & 0.9834 &  & 0.9821  &  \\ \hline
\end{tabular}
\end{table}

\subsection{Comparison and analysis of different galaxy classification models}
We replicated several classification models and used these models for galaxy classification tasks under the dataset Galaxy Zoo-The Galaxy Challenge. These models include the classical works on galaxy classification Dieleman \citep{dieleman2015rotation} and ResNet26 \citep{zhu2019galaxy}, and several typical backbone deep learning networks. The latter includes several models from the CNN series, i.e., VGG \citep{simonyan2014very}, GoogleNet \citep{szegedy2015going}, ResNet \citep{he2016deep}, EfficientNet \citep{ tan2019efficientnet} and a transformer series model called Vision Transformer \citep{dosovitskiy2020image}. To test the effectiveness of the method HIWL, we also investigated the version of replacing the backbone model of HIWL with each of the above models and performed the same classification task on the same dataset. The results of the experimental comparison are shown in Table \ref{table:scheme}.

\begin{table}
\centering
\caption{\label{table:scheme}Comparison of the test results before and after combining the method HIWL with the classical model. The table contains 11 comparison models such as Dieleman, ResNet, Vision Transoformer, etc. Avg acc represents the average accuracy of 10 runs for each model, Avg acc (with HIWL) represents the average accuracy of 10 runs for each model after incorporating HIWL, and Promotion represents the and the difference between the accuracy of the model after incorporating HIWL and the original model.}
\resizebox{\linewidth}{!}{
\huge
\begin{tabular}{cccc}
\hline
Model & Avg acc & \begin{tabular}[c]{@{}c@{}}Avg acc\\ (with HIWL)\end{tabular} & Promotion \\ \hline
Dieleman\citep{dieleman2015rotation} & 0.9337 & 0.9400 & 0.0063 \\
ResNet26\citep{zhu2019galaxy} & 0.9074 & 0.9147 & 0.0073 \\
VGG16\citep{simonyan2014very} & 0.9431 & 0.9469 & 0.0038 \\
GoogleNet\citep{szegedy2015going} & 0.9480 & 0.9507 & 0.0027 \\
ResNet34\citep{he2016deep} & 0.9507 & 0.9597 & 0.0090 \\
ResNet50\citep{he2016deep} & 0.9469 & 0.9497 & 0.0028 \\
EfficientNet-B0\citep{tan2019efficientnet} & 0.9521 & 0.9577 & 0.0056 \\
\textbf{EfficientNet-B1}\citep{tan2019efficientnet} & \textbf{0.9542} & \textbf{0.9612} & 0.0070 \\
EfficientNet-B2\citep{tan2019efficientnet} & 0.9503 & 0.9580 & 0.0077 \\
Vision Transformer\citep{dosovitskiy2020image} & 0.9264 & 0.9451 & \textbf{0.0187} \\ \hline
\end{tabular}
}
\end{table}

It is shown from Table \ref{table:scheme} that after incorporating the HIWL method, each classic model has different degrees of improvement. Among the 11 network models, the average accuracy of the EfficientNet-B1 model before and after incorporating the method HIWL is the highest. The improvement of the ViT model is the highest among the 11 models, and the reason for the higher improvement compared to other models is mainly because of the training strategies of this method. The ViT model requires a large amount of data, and our method HIWL uses the training strategy of transfer learning, which reduces ViT's demand for a large amount of data, resulting in a significant improvement in ViT on this task.

Based on the highest validation accuracy (except \citet{reza2021galaxy} and \citet{lin2021galaxy}) of each model, we compared the HIWL with galaxy classification works based on the Galaxy Zoo dataset in recent years. These available works are based on deep learning methods such as ANN, CNN, and Vision Transformer, which have been widely studied in recent years. These works are \citet{reza2021galaxy}, \citet{zhu2019galaxy}, \citet{zhang2022classifying}, \citet{gupta2022galaxy}, \citet{silva2019classificaccao},\citet{goyal2020morphological}, \citet{jimenez2020galaxy}, \citet{lin2021galaxy} and \citet{kalvankar2020galaxy}. The comparison results are presented in Table \ref{table:paper compare}. The results of the comparing methods are all extracted from the original articles.

\begin{table}
\centering
\caption{\label{table:paper compare}Comparison between the HIWL and nine other galaxy classification works based on the Galaxy Zoo dataset in literature. In this table, Overall val acc represents the highest overall accuracy on validation set, and Overall test acc represents the highest overall accuracy on test set. Num classes represents the number of classes to be divided. The accuracies of \citet{reza2021galaxy} and \citet{lin2021galaxy} are based on the test set, and the others are based on the validation set. }
\resizebox{\linewidth}{!}{
\begin{tabular}{cccc}
\hline
Method & \begin{tabular}[c]{@{}c@{}}Overall \\ val acc\end{tabular} & \begin{tabular}[c]{@{}c@{}}Overall \\ test acc\end{tabular} & Num classes \\ \hline
ANN\citet{reza2021galaxy} &  & \textbf{98.2\%} & 4 \\
ResNet26\citet{zhu2019galaxy} & 95.21\% &  & 5 \\
SC-Net\citet{zhang2022classifying} & 94.70\% &  & 5 \\
NODE-ACA\citet{gupta2022galaxy} & 95.00\% &  & 5 \\
\citet{silva2019classificaccao} & 94.01\% &  & 6 \\
layered CNN\citet{goyal2020morphological} & 88.33\% &  & 3 \\
\citet{jimenez2020galaxy} & 96.43\% &  & 2 \\
ViT\citet{lin2021galaxy} &  & 81.21\% & \textbf{8} \\
EfficientNet-B5\citet{kalvankar2020galaxy} & 93.70\% &  & 7 \\
\textbf{HIWL} & \textbf{97.22}\% & 96.32\% & 5 \\ \hline
\end{tabular}
}
\end{table}

It is shown from Table \ref{table:paper compare} that the overall accuracies of most of the models exceeds 90\%, even around 95\%. However, there are many models which are not enough attention to the minority classes. For example, ANN\citep{reza2021galaxy} studies the recognition problem of four classes: elliptical, merge, spiral, star, and then merge and star are minority classes. But the recall, precision and F1-Score of star are all lower than 0.6, while the recall, precision and F1-Score of merger are all lower than 0.2. On the whole, there exist some relationships between the number of divided classes and the final overall accuracy. Generally speaking, the larger the number of classes, the lower the overall accuracy will be. Four of the compared works in the table have a class number of 5, and our HIWL has a higher overall accuracy than the other three works. It is also shown that HIWL has the second highest overall accuracy (97.22\% and 96.32\%) after ANN \citep{reza2021galaxy}, which dealt with the classification of four classes of galaxy. It is worth noting that the ratio of the number of samples in the minority classes to the majority classes in ANN\citep{reza2021galaxy} is too small (< 1:100). Therefore, the low recognition rate in the minority classes has little impact on the final overall accuracy.

Particularly, this paper conducted some comparisons with three related works in literature based on recall, precision, and F1-Score on validation set. Each of these works studied the recognition of CRS, IBS, CSS, EO and SPI. The SC-Net \citep{zhang2022classifying} did not give the information on the precision and F1-Score. Therefore, we didn't compare the HIWL with it based on these two metrics. 

\begin{table}
\centering
\caption{\label{table:5classes compare recall}Comparison of recall between the HIWL and three literature works based on the validation set. Each of these works focuses on the recognition of CRS, IBS, CSS, EO and SPI. }
\resizebox{\linewidth}{!}{
\huge
\begin{tabular}{cccccc}
\hline
\multirow{2}{*}{Method} & \multicolumn{5}{c}{Recall} \\ \cline{2-6} 
 & CRS & IBS & CSS & EO & SPI \\ \hline
ResNet26\citep{zhu2019galaxy} & 0.9634 & 0.9431 & 0.5862 & 0.9485 & 0.9782 \\
SC-Net\citep{zhang2022classifying} & \textbf{0.9785} & \textbf{0.9785} & \textbf{0.7833} & 0.8259 & 0.9850 \\
NODE\_ACA\citep{gupta2022galaxy} & 0.9592 & 0.9425 & 0.4894 & 0.9426 & 0.9268 \\
\textbf{HIWL} & 0.9715 & 0.9727 & 0.7414 & \textbf{0.9718} & \textbf{0.9897} \\ \hline
\end{tabular}
}
\end{table}

\begin{table}
\centering
\caption{\label{table:5classes compare precision}Comparison of precision between the HIWL and two literature works based on the validation set. Each of these works focuses on the recognition of CRS, IBS, CSS, EO and SPI. }
\resizebox{\linewidth}{!}{
\huge
\begin{tabular}{cccccc}
\hline
\multirow{2}{*}{Method} & \multicolumn{5}{c}{Precision} \\ \cline{2-6} 
 & CRS & IBS & CSS & EO & SPI \\ \hline
ResNet26\citep{zhu2019galaxy} & 0.9611 & 0.9561 & 0.7234 & 0.9412 & 0.9573 \\
NODE\_ACA\citep{gupta2022galaxy} & 0.9621 & 0.9001 & 0.6053 & 0.9048 & 0.9565 \\
\textbf{HIWL} & \textbf{0.9808} & \textbf{0.9632} & \textbf{0.7963} & \textbf{0.9668} & \textbf{0.9872} \\ \hline
\end{tabular}
}
\end{table}

\begin{table}
\centering
\caption{\label{table:5classes compare F1-Score}Comparison of F1-Score between the HIWL and two literature works based on the validation set. Each of these works focuses on the recognition of CRS, IBS, CSS, EO and SPI. }
\resizebox{\linewidth}{!}{
\huge
\begin{tabular}{cccccc}
\hline
\multirow{2}{*}{Method} & \multicolumn{5}{c}{F1-Score} \\ \cline{2-6} 
 & CRS & IBS & CSS & EO & SPI \\ \hline
ResNet26\citep{zhu2019galaxy} & 0.9622 & 0.9495 & 0.6476 & 0.9448 & 0.9677 \\
NODE\_ACA\citep{gupta2022galaxy} & 0.9607 & 0.9208 & 0.5412 & 0.9233 & 0.9414 \\
\textbf{HIWL} & \textbf{0.9762} & \textbf{0.9679} & \textbf{0.7679} & \textbf{0.9693} & \textbf{0.9885} \\ \hline
\end{tabular}
}
\end{table}

From Table \ref{table:5classes compare recall}, it is shown that although SC-Net \citep{zhang2022classifying} has a higher recall than HIWL on CRS, IBS and CSS, it has a significantly lower recall on EO (>10\%) than the other three models. In addition, the recall of HIWL in each class is higher than that of ResNet26 \citep{zhu2019galaxy} and NODE-ACA \citep{gupta2022galaxy}. In the the comparisons based on precision (Table \ref{table:5classes compare precision}), the proposed HIWL obtained the highest performance on every class. In the comparison based on F1-Score (Table \ref{table:5classes compare F1-Score}), the HIWL still achieves the highest performance. On the whole, therefore, the proposed HIWL is superior to the other typical works based on three metrics (recall, precision, F1-Score).

\subsection{Model Feature Visualization}

To observe the characteristics of information extraction of the model HIWL, this work performed the visualization of the features extracted by the models. The output of the model's penultimate layer is taken as the extracted features. The models we investigated are EfficientNet-B1 and HIWL with EfficientNet-B1 as the backbone. The experimental studies in Table \ref{table:scheme} show that these two models have the best performance. Each model is visualized on training samples and test samples. 
To ensure the observability, only 5,000 galaxy samples randomly selected from them were visualized. There are 2,881 samples in the test set, and they all were used in visualization.
 
\begin{figure*}
 \subfigure[EfficientNet-B1]{\includegraphics[width=0.36\linewidth]{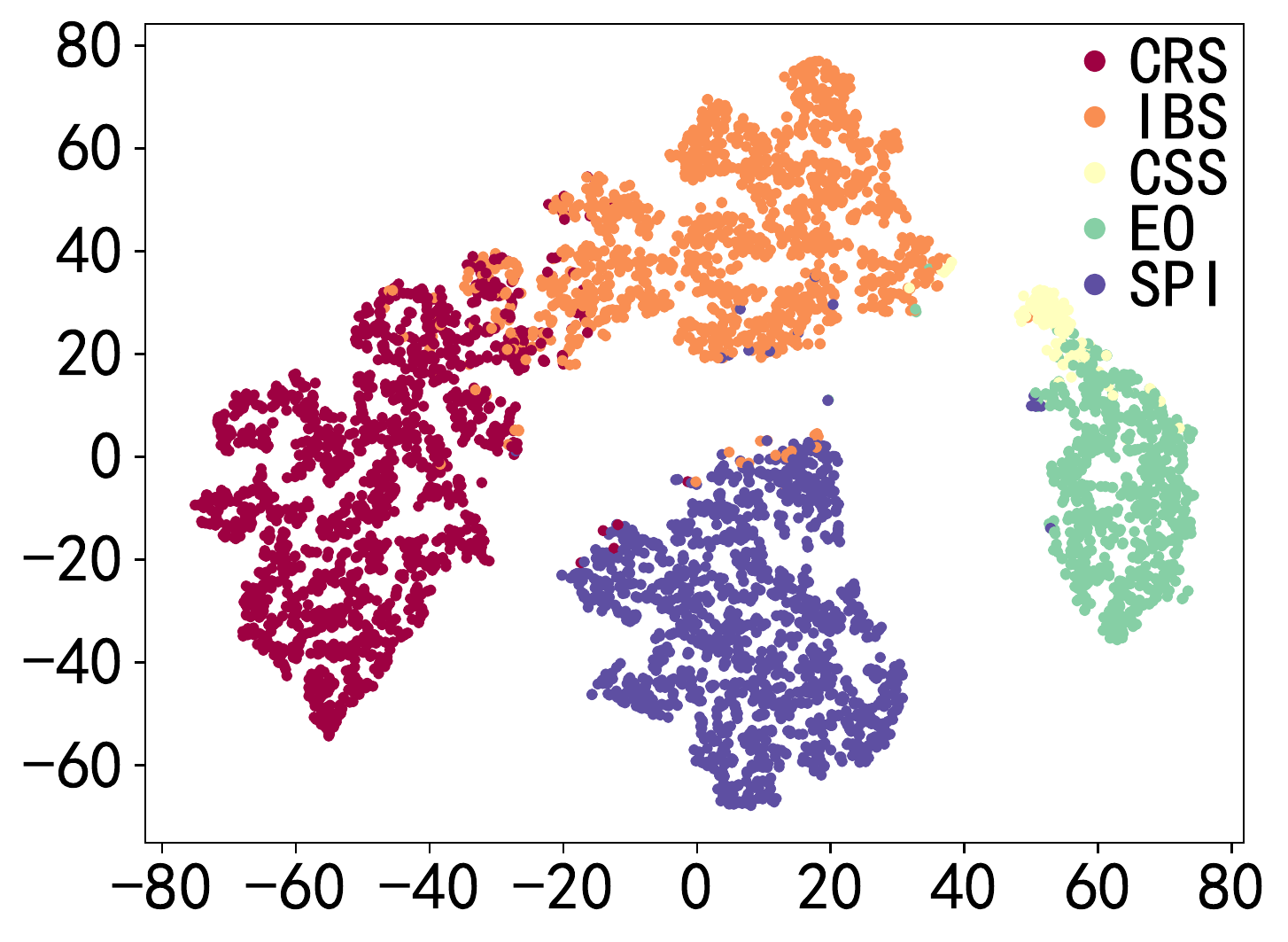}\label{subfig:fc_noscheme_train}}
 \subfigure[HIWL]{\includegraphics[width=0.36\linewidth]{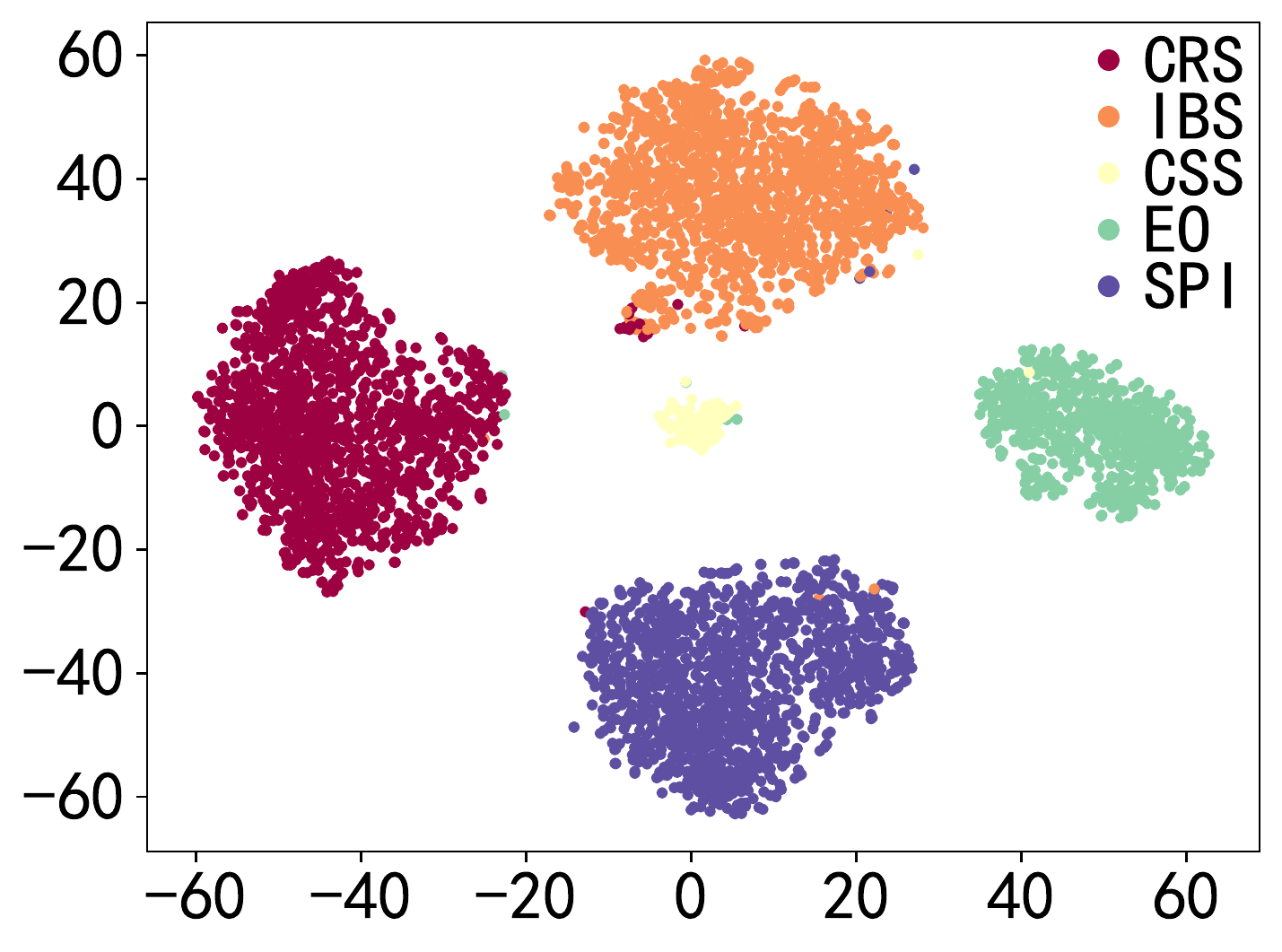}\label{subfig:fc_scheme_train}} \\
 \subfigure[EfficientNet-B1]{\includegraphics[width=0.36\linewidth]{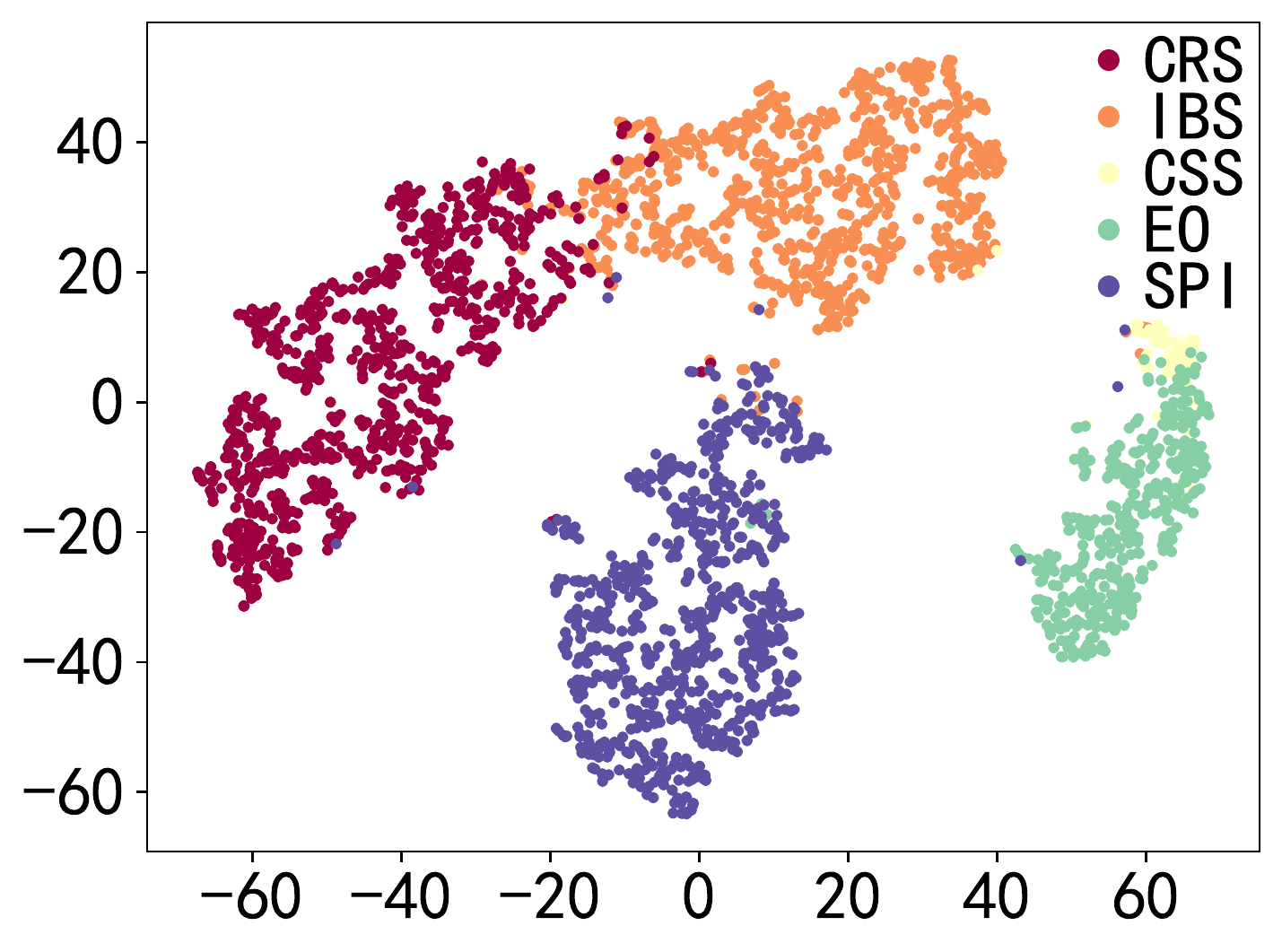}\label{subfig:fc_noscheme_test}}
 \subfigure[HIWL]{\includegraphics[width=0.36\linewidth]{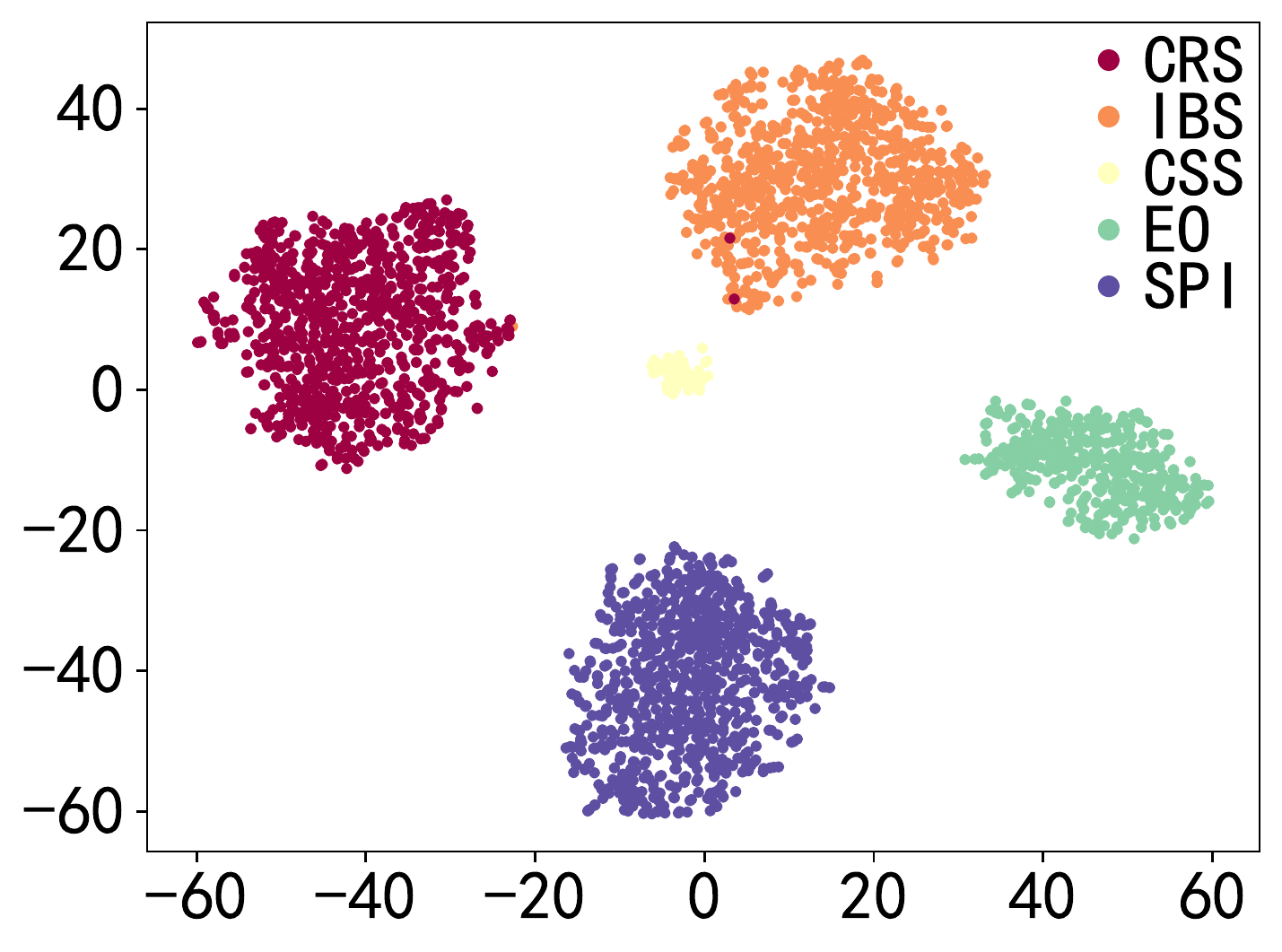}\label{subfig:fc_scheme_test}}
  \caption{\label{fig:fc_visual}
  Visualization of model features. (a) and (b) present the extracted features of 5000 randomly selected training samples, (c) and (d) the extracted features of test samples. It is shown that the features of HIWL have more discriminated information than those of EfficientNet-B1.
  }
\end{figure*}

For the EfficientNet-B1 model, we directly input the selected training data or the test data into the model. 
After obtaining the output of the penultimate layer, the operations of flattening, dimensionality reduction, and visualization are performed sequentially. 
For HIWL, we did the following operations: (1) The samples labeled as CRS, IBS, and SPI are sent into the four-classification model of HIWL, and the output features of the penultimate layer are obtained; (2) the samples labelled CSS and EO are fed into the two-classification model to obtain the output features of the penultimate layer; (3) the features obtained from these samples are flattened and projected into a 2-dimensional space using t-SNE.
The result is shown in Figure \ref{fig:fc_visual}.
 
From Figure \ref{fig:fc_visual}, it is shown that the visualization results of EfficientNet-B1 or HIWL are consistent in the training data and test data respectively. In the feature space of EfficientNet-B1, the sample points within the class are relatively sparse; the distance of sample points between different classes is small and entangled with each other, resulting in inconspicuous class boundaries.
 Contrastingly, in the feature space of HIWL, the sample points within the class become denser; the distance of sample points between the classes becomes larger and the phenomenon of intertwining is greatly reduced, resulting in obvious class boundaries. Although multiple techniques in the method HIWL contribute to this change, more contributions come from label smoothing, which is also consistent with \citet{Mller2019WhenDL}. These Characteristics of HIWL features are beneficial to alleviating the problem of model misclassification and making the classification more accurate. 

In addition, it is shown that there is a certain similarity between CRS and IBS, and the similarity between EO and CSS is high. This similarity is demonstrated by the fact that the samples between them partially overlap in feature space. That is to say, some of the samples of EO are partially in the space where CSS is located, and vice versa.

\subsection{Model Attention Visualization}
\begin{figure*}
  \centering
  \includegraphics[width=16cm]{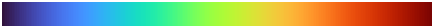}\label{color}
  \subfigure[CRS]{\includegraphics[width=3.2cm]{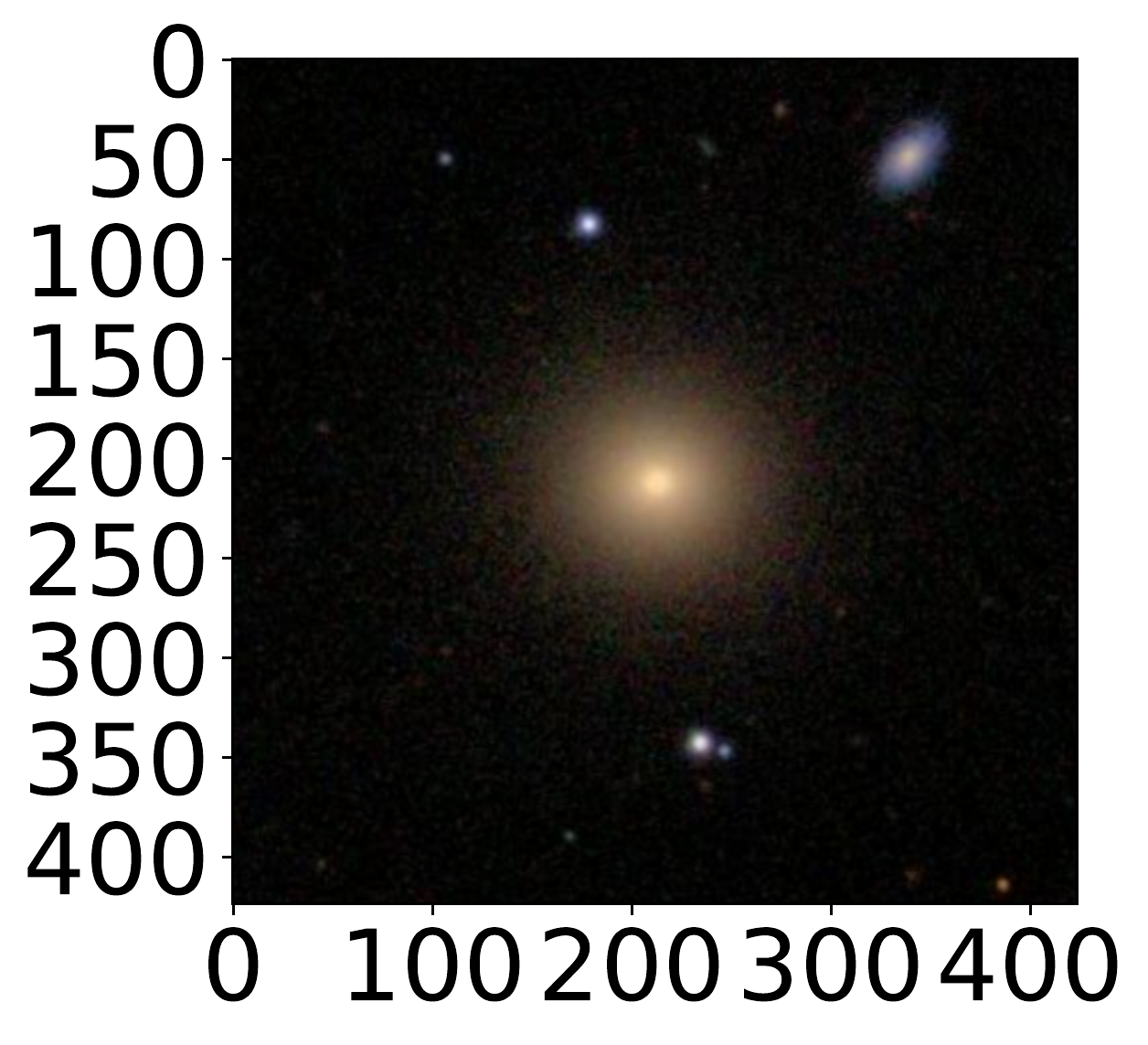}\label{grad_cam_origin_3}}
  \subfigure[IBS]{\includegraphics[width=3.2cm]{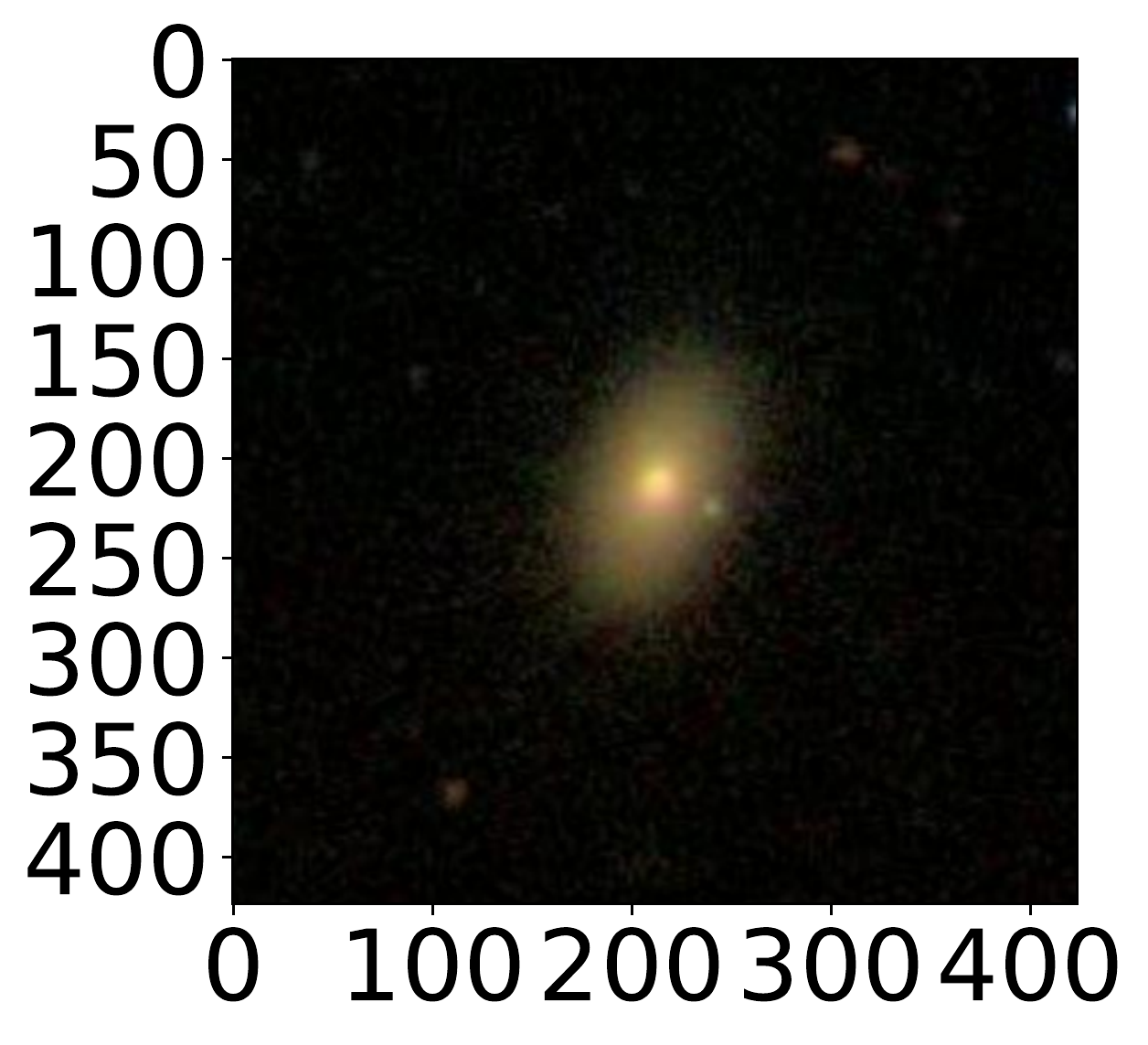}\label{grad_cam_origin_0}}
  \subfigure[CSS]{\includegraphics[width=3.2cm]{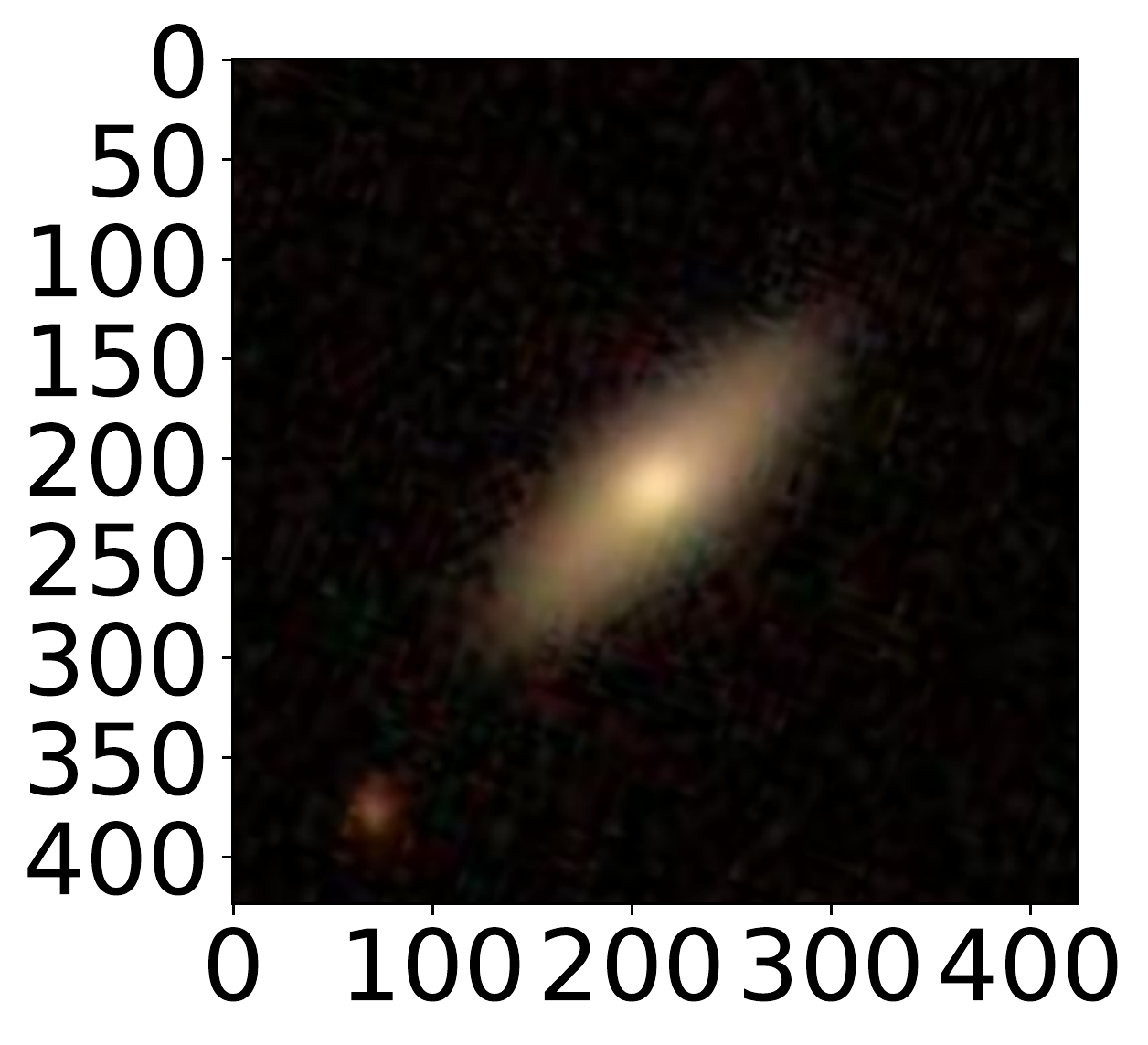}\label{grad_cam_origin_1}}
  \subfigure[EO]{\includegraphics[width=3.2cm]{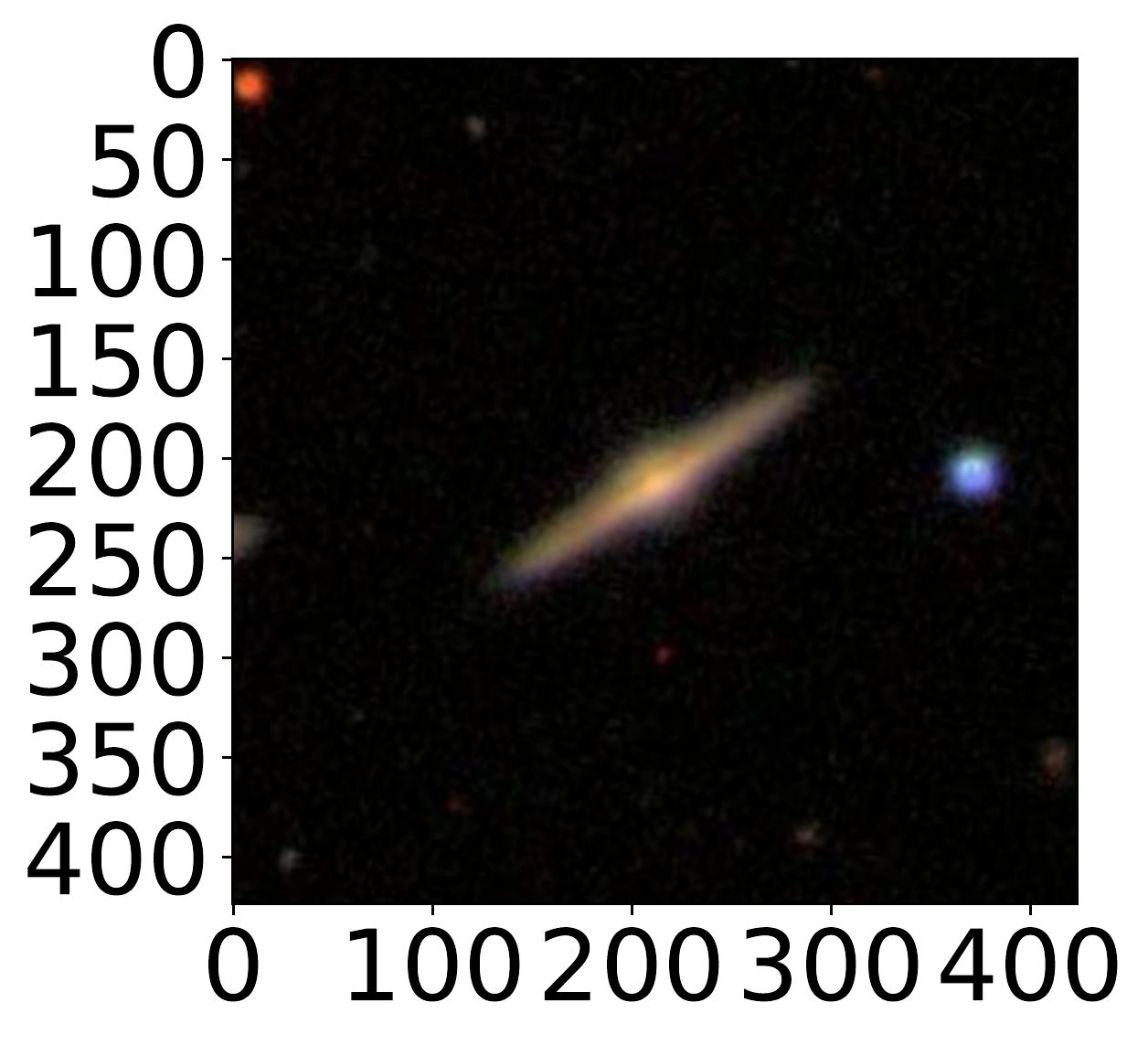}\label{grad_cam_origin_2}}
  \subfigure[SPI]{\includegraphics[width=3.2cm]{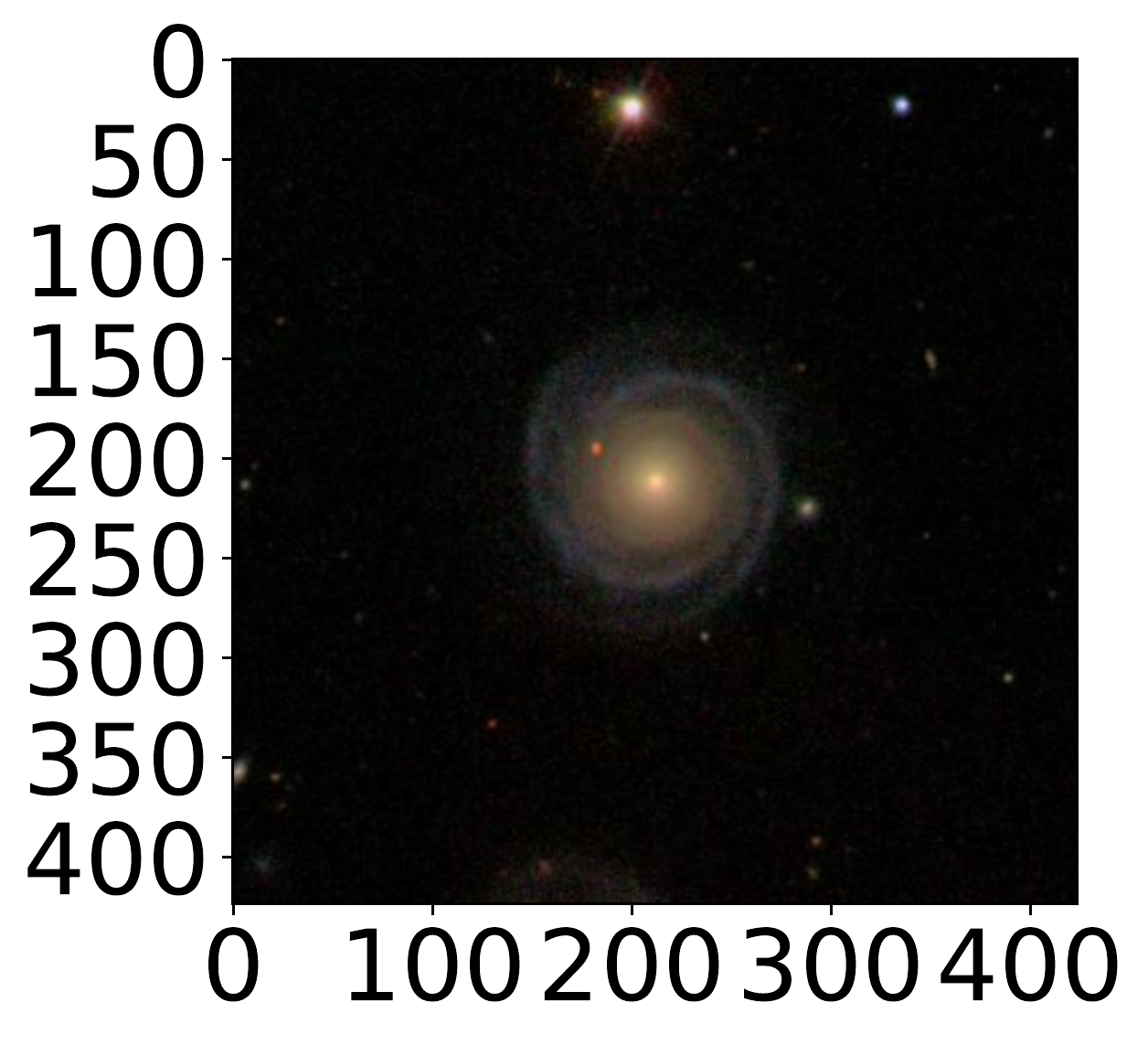}\label{grad_cam_origin_4}}
  
  \subfigure[EfficientNet-B1]{
  \hspace{-0.1cm}\includegraphics[width=3.2cm]{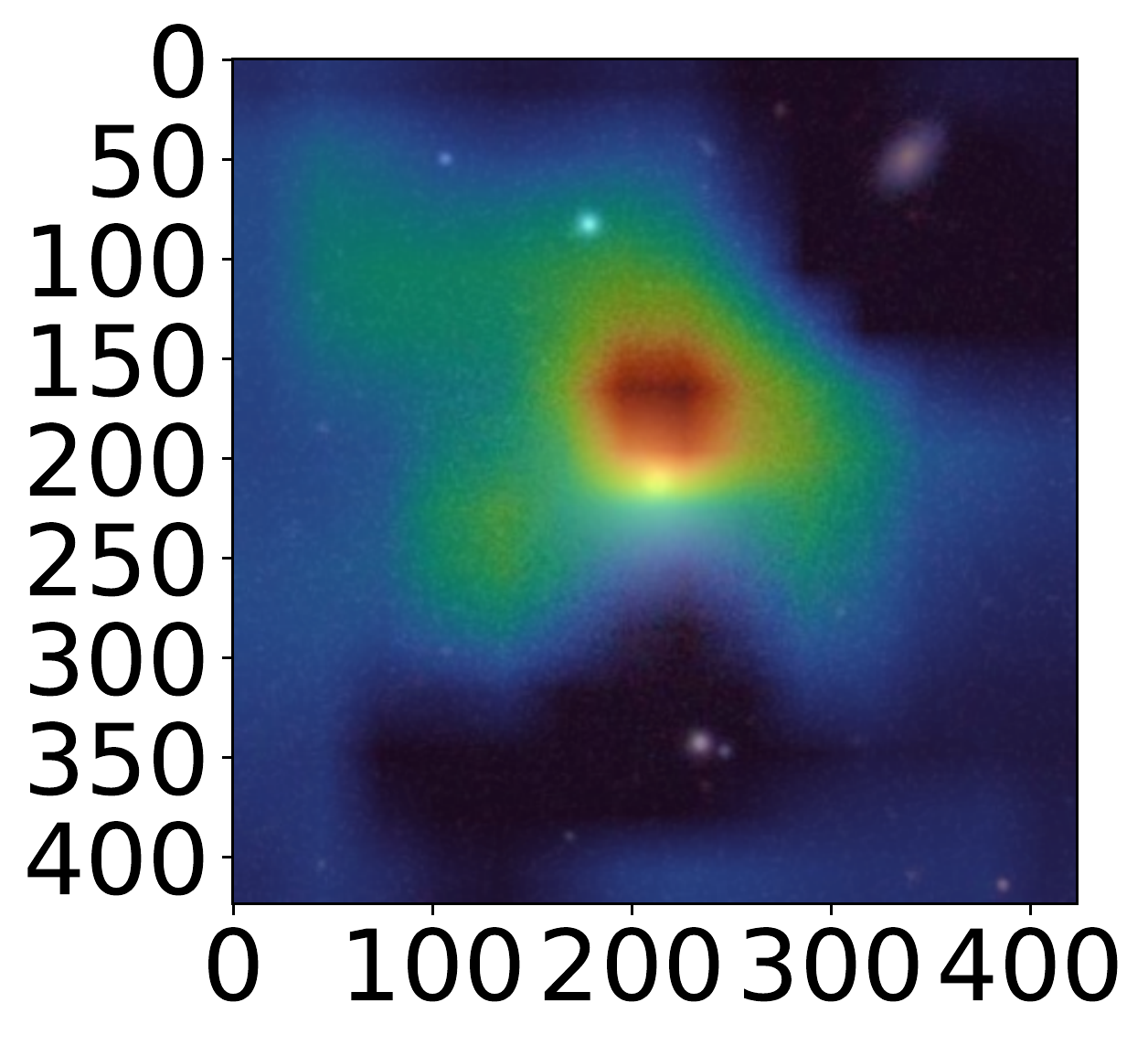}\label{grad_cam_noscheme_3}
  \includegraphics[width=3.2cm]{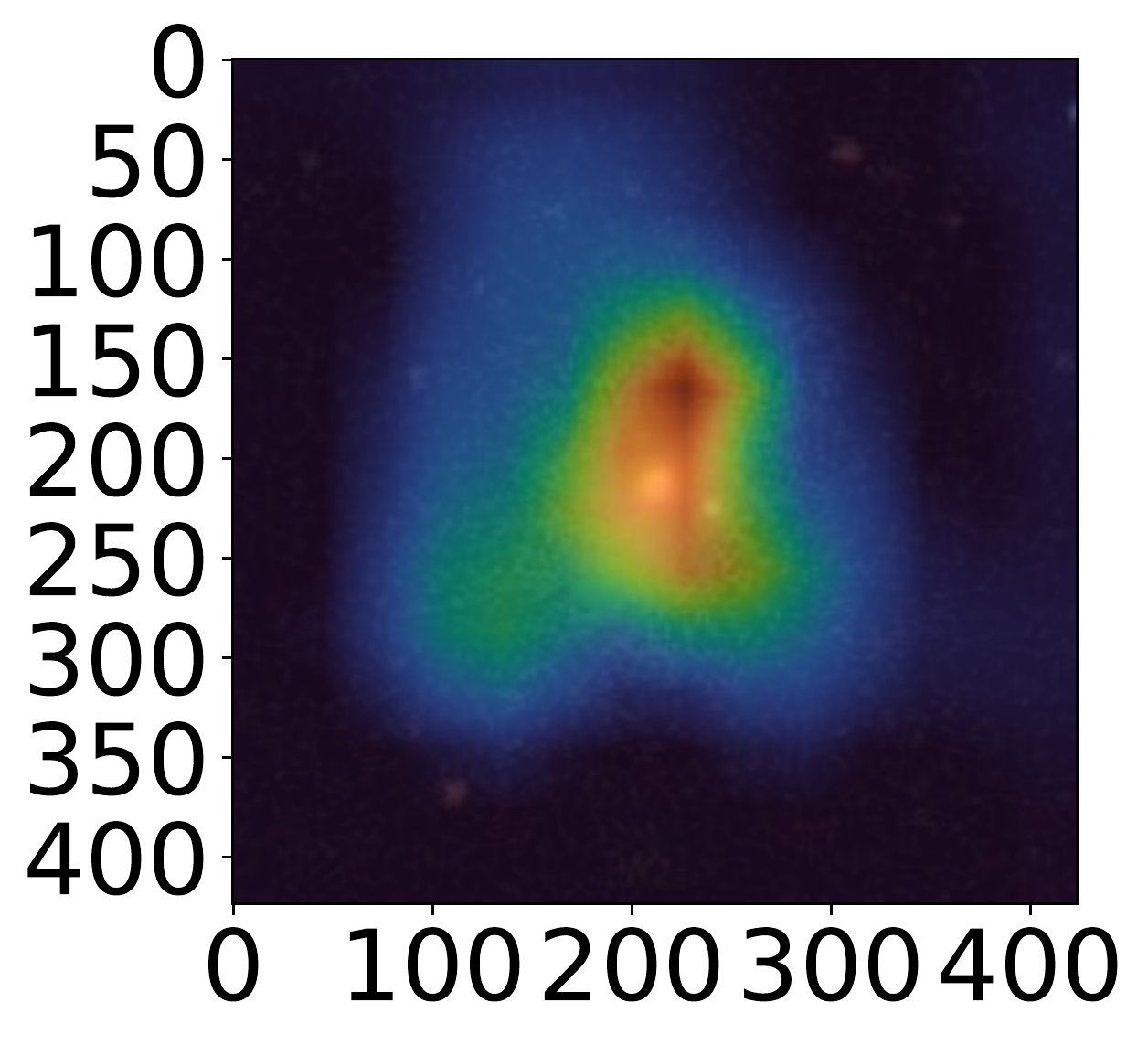}\label{grad_cam_noscheme_0}
  \includegraphics[width=3.2cm]{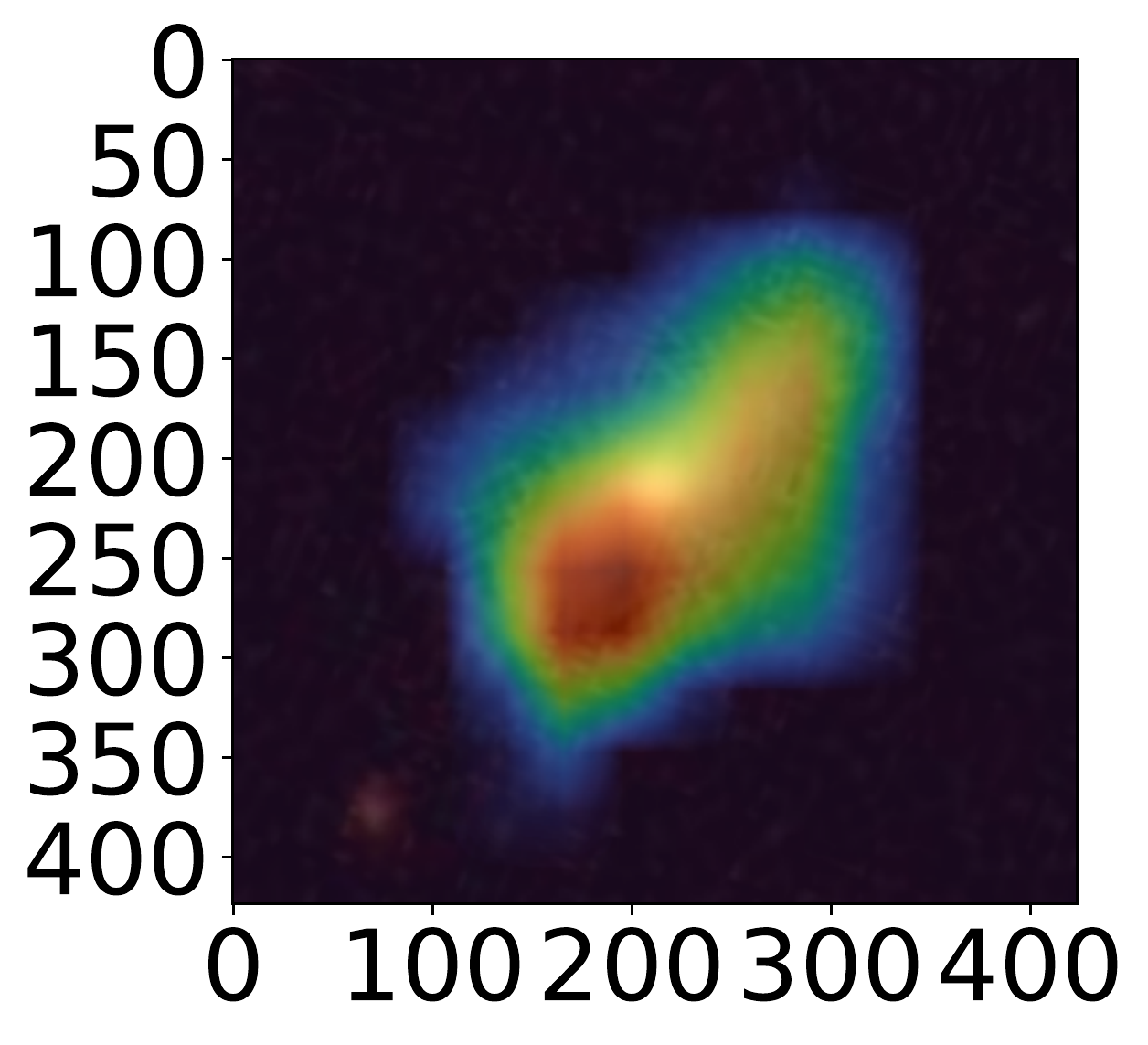}\label{grad_cam_noscheme_1}
  \includegraphics[width=3.2cm]{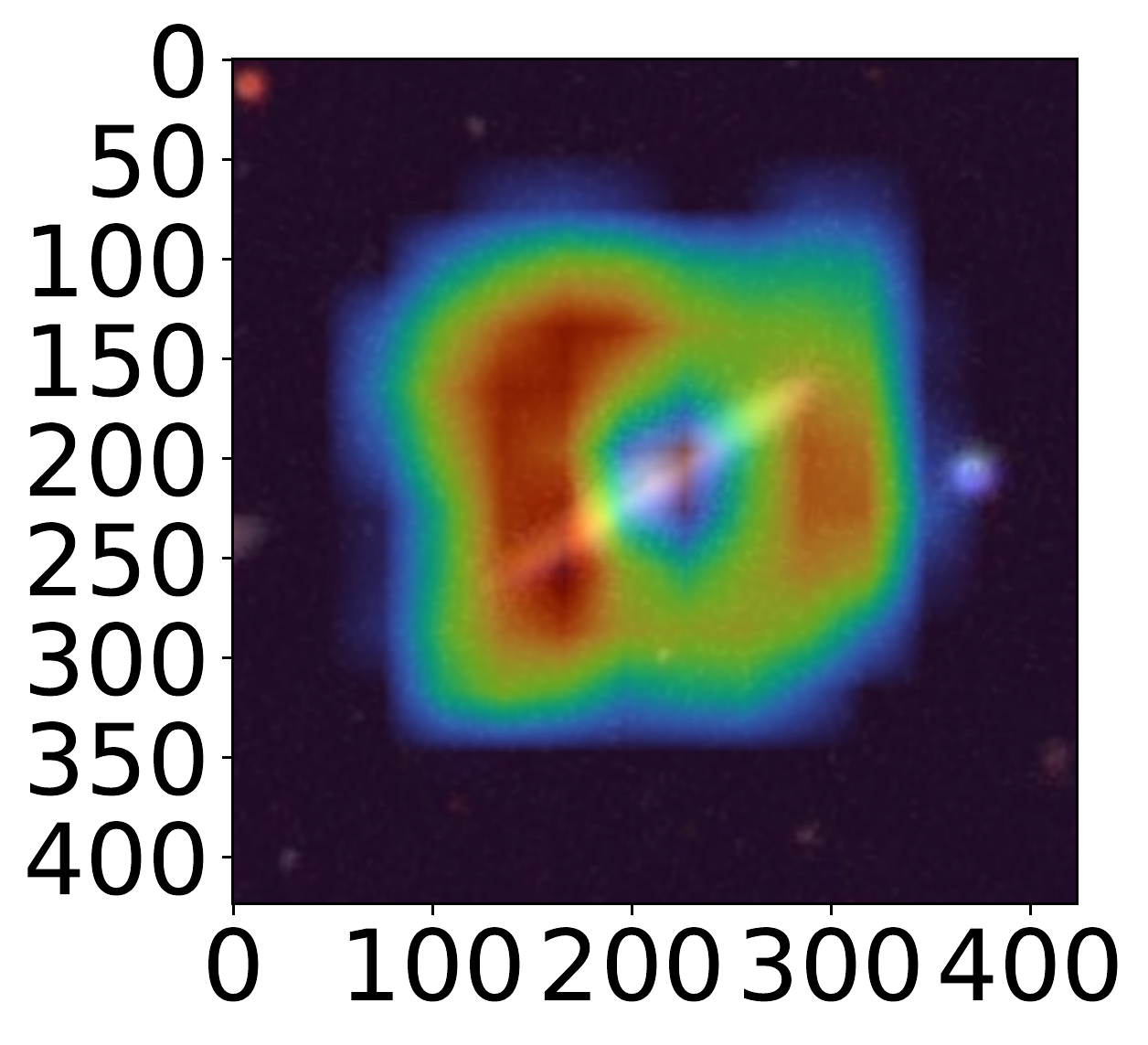}\label{grad_cam_noscheme_2}
  \includegraphics[width=3.2cm]{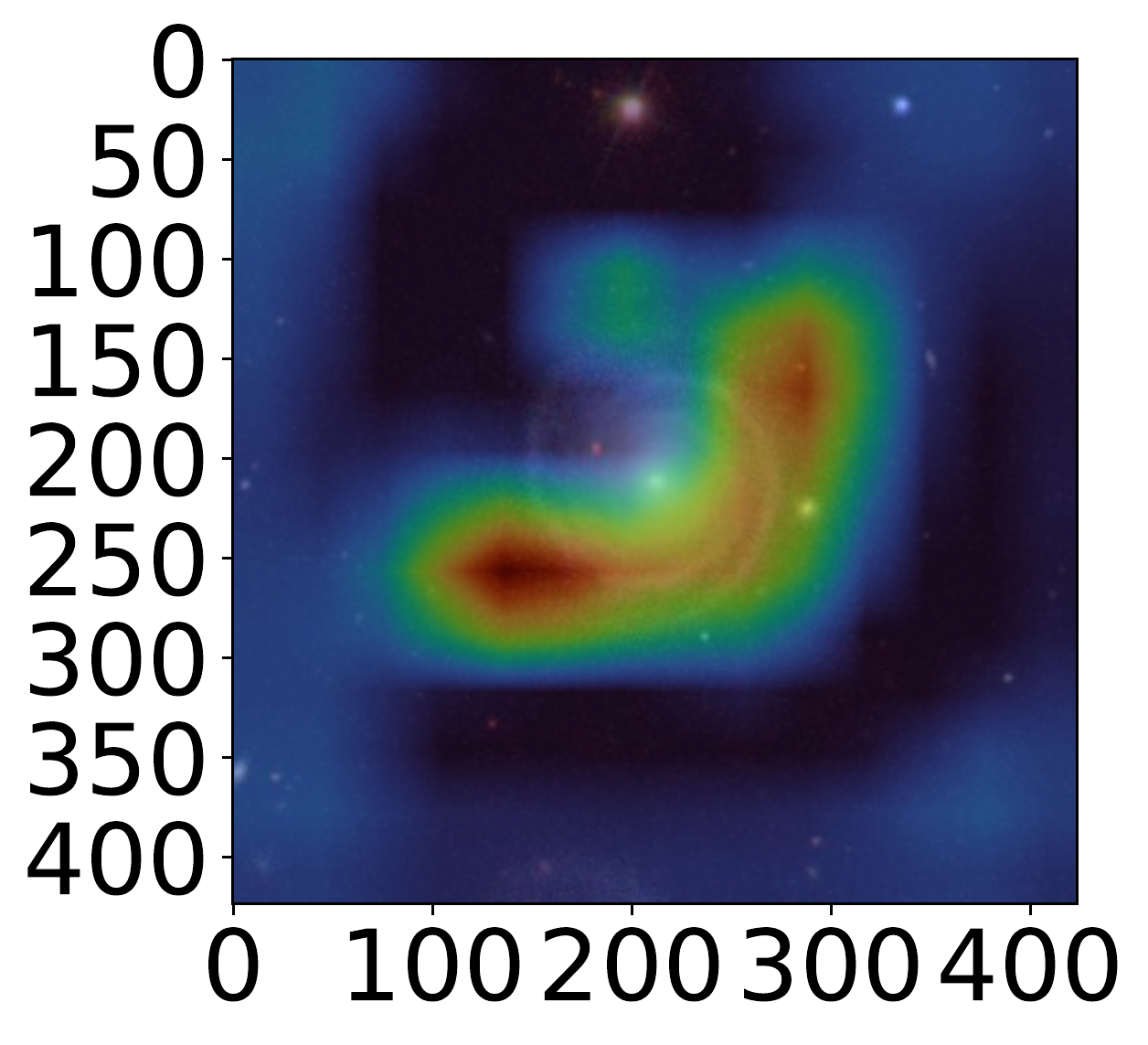}\label{grad_cam_noscheme_4}}
  
  \subfigure[HIWL]{
  \hspace{-0.02cm}\includegraphics[width=3.2cm]{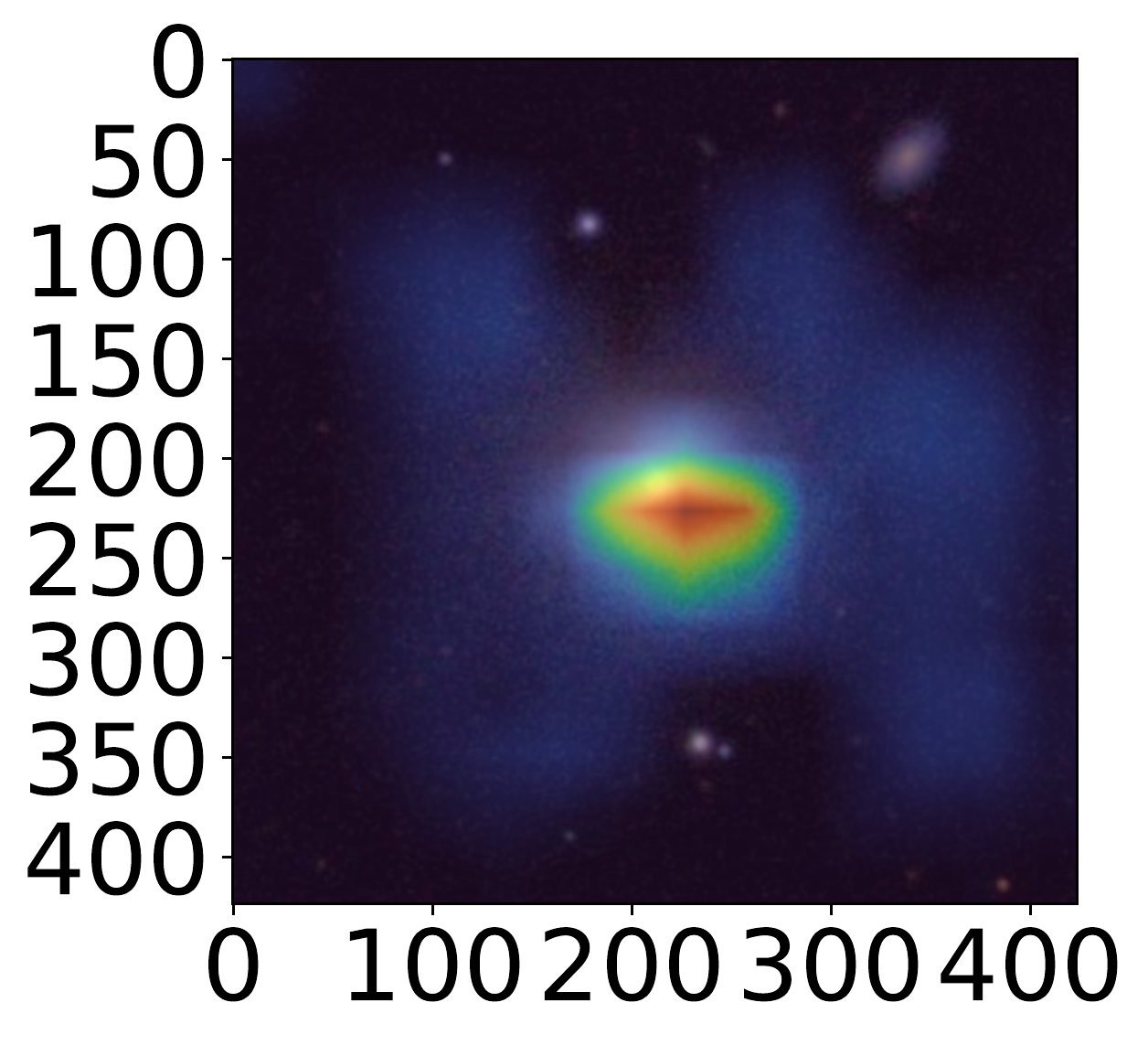}\label{grad_cam_scheme_3}
  \includegraphics[width=3.2cm]{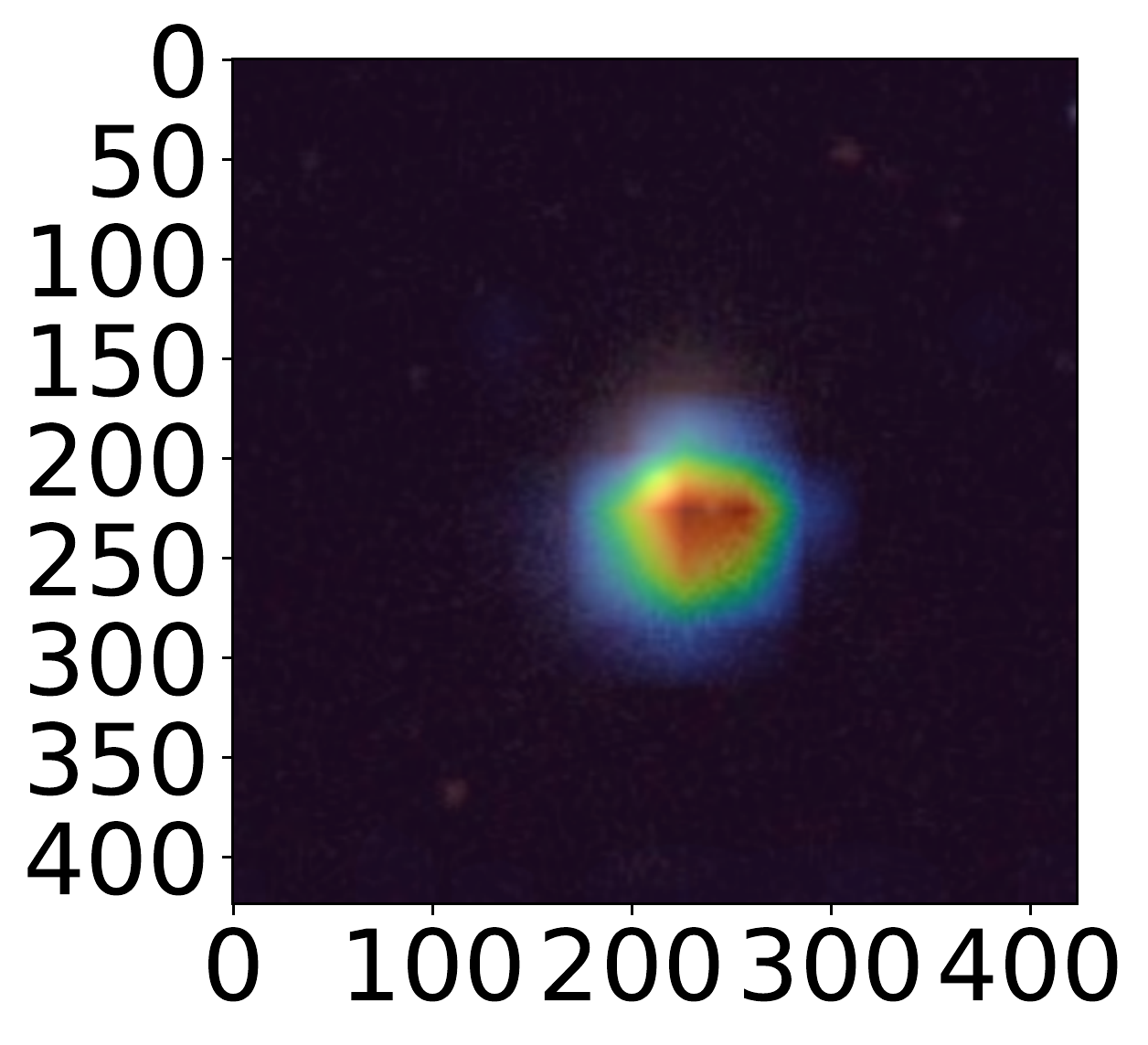}\label{grad_cam_scheme_0}
  \includegraphics[width=3.2cm]{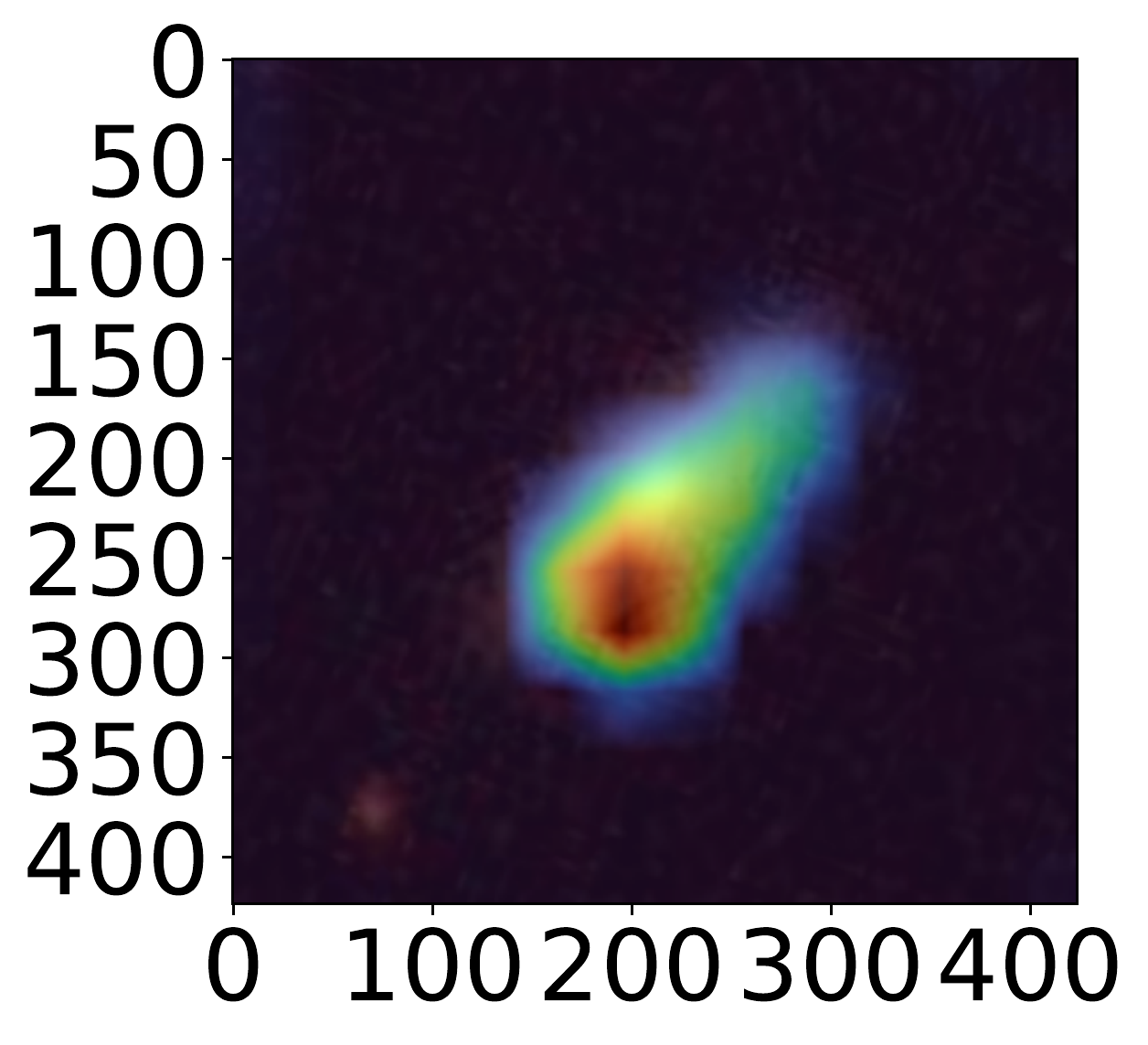}\label{grad_cam_scheme_1}
  \includegraphics[width=3.2cm]{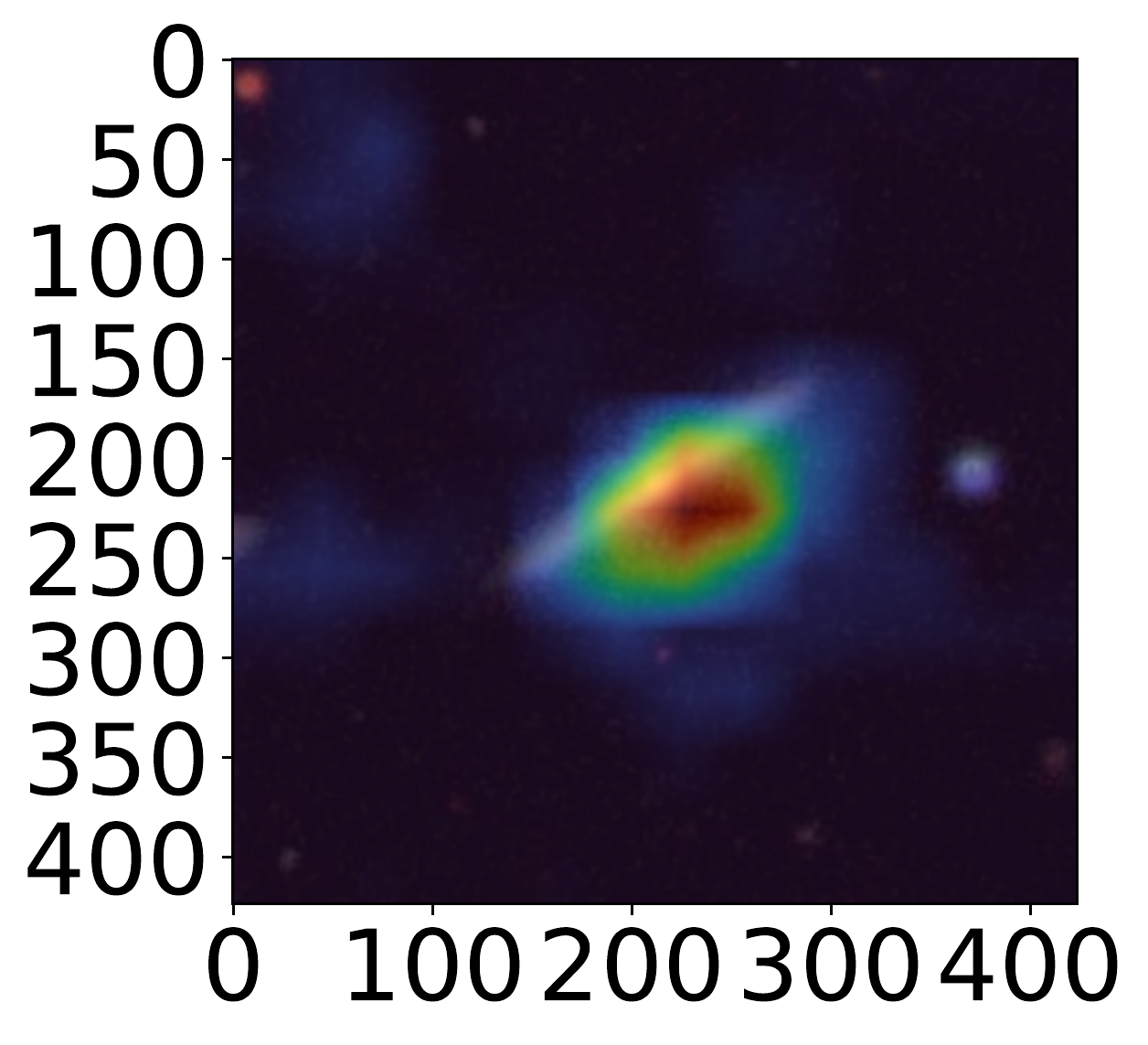}\label{grad_cam_scheme_2}
  \includegraphics[width=3.2cm]{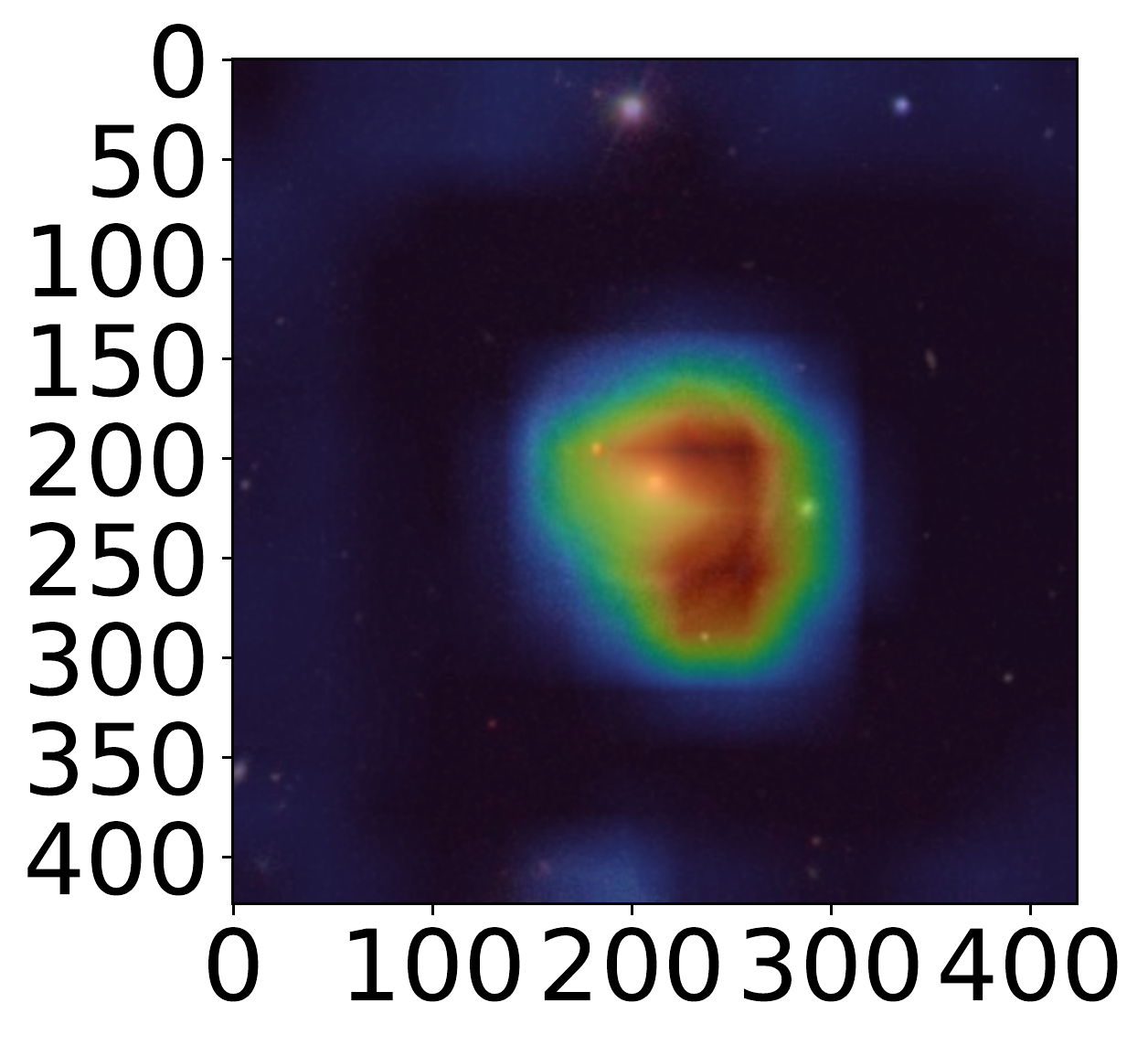}\label{grad_cam_scheme_4}}
  \caption{\label{fig:grad_cam_visual}Visualisation using GRAD-CAM method: the more red it is, the larger the value is, and the more attention the region is in the GRAD-CAM visualization. The first row represents the five original galaxy samples belonging to different classes; the second row represents the GRAD-CAM visualisation of EfficientNet-B1; the third row represents the GRAD-CAM visualisation of HIWL.}
\end{figure*}

For the general deep learning network, many people consider it to be a black box with weak interpretability. For example, why it predicts as it does and where it focuses are unknown. Grad-CAM (Gradient-weighted Class Activation Mapping) \citep{selvaraju2017grad} is a deep network visualization method based on gradient localization. It uses the gradient of the target (such as the logits of a class in a classification task), which flows into the final convolutional layer, to generate a rough localization map to highlight important regions in the image for prediction. It explains the classification basis of the deep neural network model in the form of a heat map, which facilitates humans to understand the model and analyse it. In order to understand the image regions that the model focuses on when classifying galaxies, we then visualised the heat map for each of the five classes using GRAD-CAM, and the results are shown in Figure \ref{fig:grad_cam_visual}. In this visualization, we compared two models, EfficientNet-B1 and HIWL. In this figure, the closer to red, the more the model pays attention to this region, and the closer to black, the less the model pays attention to this region.

In general, the regions to be focused on are larger and scattered when using EfficientNet-B1 for galaxy image recognition, whereas it is smaller and aggregated for HIWL. Both models are not affected by small bright spots (other galaxies) in the image and still focus on the main regions of the class to be identified. This robustness benefits from the high noise immunity of the EfficientNet series of models. In particular, for CRS, the method HIWL focuses on the middle and edge contours, while EfficientNet-B1 also focuses on a few redundant surrounding pixels. For IBS, both models focus on the aspect ratio and small surrounding regions, but EfficientNet-B1 focuses on more scattered regions. For CSS, both models have similar focus regions and the difference between them is that the focus regions of HIWL are more aggregated. For EO, EfficientNet-B1 focuses on large and dispersed regions containing both tips of the EO, as well as the side surrounding areas, while HIWL focuses on the central width and surrounds which are relatively more aggregation. For SPI, EfficientNet-B1 focuses on less spiral arms regions and more dispersed surrounding regions, while HIWL focuses on the spiral arms. Therefore, the focus of HIWL are excellent and compact, while the attention of EfficientNet-B1 are drawn to some regions with redundancies.

\section{Conclusion}
This work carries out a series of research to address several typical problems in deep learning-based methods for classifying galaxy images. For example, the auto learning of galaxy features, the masking effect due to the different degree of similarities between classes, the class imbalance problem, and the machine learning problems caused by the discrepancy between the discrete representation of Galaxy classes and the essentially gradual changing of morphology (DDRGC). For the auto learning of galaxy features, this paper explores the application of the EfficientNet model, which combines both representational power and efficiency in galaxy information extraction. To deal with the masking effects mentioned above and the imbalance effects, we utilized the idea of hierarchical learning and weighted sampling + online data augmentation techniques. 
For the problems caused by the DDRGC discrepancy, this paper explores label smoothing techniques to reduce this discrepancy and its negative effects.

The proposed scheme was evaluated on the Galaxy Zoo-The Galaxy Challenge data.
 This problem is to deal with the classification of five types of galaxy images: CRS, IBS, CSS, EO, and SPI. 
It is shown that the proposed scheme achieves an overall classification accuracy of 96.32\%. It still has a certain improvement when the idea of this method is applied to other typical classification models. In further comparisons with recent works on galaxy classification, the proposed HIWL shows advantages over three evaluation metrics (recall, precision, and F1-Score). In particular, the method shows a significant improvement in the recognition of CSS, with a recall of 70.69\% and a precision of 78.55\%. To understand the recognition effectiveness and basis of the model, we also use t-SNE to visualize model features, and use GRAD-CAM method to visualize interest area of the model as a heatmap.

The typical characteristic of HIWL is the improvement of the recognition effect on the images of the minority class of galaxies. In fact, the rare and special celestial bodies are of special value in astronomical research. Therefore, a very typical potentially application value of the method in this paper is the application scenario of astronomical data processing in the search for such rare and special celestial bodies. In addition, although the recognition based on galaxy images in this paper verifies the effectiveness of the method. In fact, the core techniques of HIWL are hierarchical classification, weighted sampling and label smoothing. These techniques can be applied not only to the recognition of galaxy images, but also to a wider range of astronomical data processing problems such as other types of astronomical images, spectral and time-series observational data.

In future work, we will combine label smoothing with morphological features in a more nuanced way. That is, the smooth value can reflect the similarity between classes, so that classes with different degrees of similarity have different and meaningful values on the label. In addition, we plan to explore the galaxy recognition problem on larger datasets and in more fine-grained class divisions.

\section*{Acknowledgements}

This work were supported by the National Natural Science Foundation of China (Grant No. 11973022), the Natural Science Foundation of Guangdong Province (No. 2020A1515010710), the Major projects of the joint fund of Guangdong and the National Natural Science Foundation (Grant No. U1811464).

\section*{Data and code Availability}

The dataset which named Galaxy Zoo-The Galaxy Challenge is available at \label{dataset available}\url{https://www.kaggle.com/competitions/galaxy-zoo-the-galaxy-challenge/data}. And the code for the whole experiment process for this dataset, including clean sample selection, clean sample set partitioning, pre-processing, training models and comparing models, are available at \url{https://github.com/xrli/HIWL}.


\section*{Footnotes}

software: Numpy \citep{harris2020array}, Matplotlib \citep{hunter2007matplotlib}, PyTorch \citep{paszke2019pytorch}, TorchVision \citep{marcel2010torchvision}, tqdm \citep{da2019tqdm}.



\bibliographystyle{mnras}
\bibliography{cites} 







\bsp	
\label{lastpage}
\end{document}